\documentclass[11pt]{article} 
\usepackage[utf8]{inputenc}
\usepackage{multirow}
\usepackage{amsfonts}
\usepackage{amsmath}
\usepackage{amssymb}
\usepackage{amsthm}
\usepackage{mathtools}
\usepackage{dsdshorthand}
\usepackage{tikz}
\usepackage{tensor}
\usetikzlibrary{decorations}
\usetikzlibrary{trees}
\usetikzlibrary{decorations.pathmorphing}
\usetikzlibrary{decorations.markings}
\usetikzlibrary{external}
\usetikzlibrary{intersections}
\usetikzlibrary{shapes,arrows}
\usetikzlibrary{arrows.meta}
\usetikzlibrary{calc}
\usetikzlibrary{shapes.misc}
\usetikzlibrary{decorations.text}
\usetikzlibrary{backgrounds}
\usetikzlibrary{fadings}
\usetikzlibrary{tikzmark,calc,arrows,shapes,decorations.pathreplacing}
\tikzset{
	cross/.style={cross out, draw=black, minimum size=2*(#1-\pgflinewidth), inner sep=0pt, outer sep=0pt},
	branchCut/.style={postaction={decorate},
		snake=zigzag,
		decoration = {snake=zigzag,segment length = 2mm, amplitude = 2mm}	
}}
\usepackage{caption}
\usepackage{color}
\definecolor{darkgreen}{rgb}{0,0.5,0}
\definecolor{darkblue}{rgb}{0,0,0.6}
\definecolor{purple}{rgb}{0.4,.2,0.7}
\usepackage[margin = 2.5cm]{geometry}
\usepackage{graphicx}
\usepackage[hyperindex,breaklinks,hyperfootnotes = false, colorlinks = true, linkcolor = darkblue, citecolor = purple]{hyperref}
\usepackage{subcaption}
\usepackage{ytableau}
\usepackage[english]{babel}
\usepackage[autostyle, english = american]{csquotes}
\MakeOuterQuote{"}

\usepackage{comment}

	\newcommand{\ee}{\end{equation}}
\newcommand{\bea}{\begin{eqnarray}}
	\newcommand{\eea}{\end{eqnarray}}

\def\nref#1{(\ref{#1})}

\def\nref#1{(\ref{#1})}
\def\n{\nabla}

\def\fv{\text{fv}}
\def\tv{\text{tv}}

\begin{document}
	
	\thispagestyle{empty}
	\begin{center}
		~\vspace{5mm}
		
		\vskip 2cm 
		
		{\LARGE \bf 
		  One-loop aspects of Coleman de Luccia instantons at small backreaction 
		}

		\vspace{0.5in}

		Victor Ivo$^1$,

		\vspace{0.5in}

		$^1$
		{\it  Jadwin Hall, Princeton University,  Princeton, NJ 08540, USA }

        \par\vspace{0.5em}
{\small \texttt{vi6083@princeton.edu}}
		
	\end{center}
	
	\vspace{0.5in}
	
	\begin{abstract}
        We discuss the Euclidean path integral around Coleman de Luccia instantons. We compute their contribution at the one-loop level, at leading order in the small backreaction limit. At this level of approximation, their contribution factorizes into a pure gravity and a pure matter contribution. In deriving this result, we also clarify what happens to some zero mode contributions to the path integral, once the symmetry responsible for them is broken. With these results established, we propose a formula for the decay rate of the false vacuum in terms of gravitational path integrals, and we show that our definition matches the usual quantum field theory result as $G_{N}\rightarrow0$. Lastly, we propose a prescription to study how the phase of the gravity+matter path integral changes as we change the parameters of the theory.
        \end{abstract}
	
	\vspace{1in}
	
	\pagebreak
	
	\setcounter{tocdepth}{3}
	{\hypersetup{linkcolor=black}\tableofcontents}

\section{Introduction}

The Euclidean gravitational path integral is a useful tool in understanding semiclassical aspects of gravity. This is emphasized in particular by modern developments in the black hole information paradox \cite{Almheiri:2019qdq, Penington:2019kki}, and by the investigation of Schwarzian corrections to near extremal black hole entropy \cite{Iliesiu:2020qvm, Heydeman:2020hhw}. Moreover, the Euclidean path integral is also useful in AdS/CFT \cite{Maldacena:1997re,Gubser:1998bc,Witten:1998qj}, where one can compute CFT observables, such as partition functions \cite{Gubser:1998bc,Witten:1998qj} or entanglement entropies \cite{Lewkowycz:2013nqa}, from Euclidean gravity saddles. 

However, while successful in many examples, the Euclidean gravitational path integral has a pathology commonly known as a "conformal factor problem"\cite{Gibbons:1978ac}. That is, the quadratic action around the saddles of the path integral has infinitely many negative directions. One of the perhaps oldest proposed resolutions to this problem \cite{Gibbons:1978ac, Hawking:2010nzr,Polchinski:1988ua} involves noting that this infinite number of negative directions can generally be tracked back to one field\footnote{In some gauges, this problematic field is the overall conformal factor of the metric, thus the name "conformal factor problem". The problematic field, however, is gauge fixing dependent, see \cite{Ivo:2025yek,Shi:2025amq}.}. One can therefore regulate the path integral by rotating the contour of this field to the imaginary axis \footnote{Note this does not fix other, IR related, issues of the Euclidean path integral \cite{Horowitz:2025zpx}}. This step is just a local field redefinition. In general, however, a finite number of modes in the decomposition of this field are positive and need to be "rotated back" to the real axis. Because this second step involves only a finite number of field modes, it gives an overall phase to the path integral. See, for example, \cite{Polchinski:1988ua, Hawking:2010nzr,Ivo:2025yek,Shi:2025amq}. 

Perhaps the most famous example of this phenomenon is that the sphere partition function in Euclidean gravity, $Z_{\text{GR}}(S^{D})$, has a phase, as discussed by Polchinski in \cite{Polchinski:1988ua}. To pick the phase unambiguously, one has to make the Planck constant of the theory slightly complex, as stressed in \cite{Maldacena:2024spf, Ivo:2025yek}, and depending on which way we complexify it, we obtain
\begin{equation}
\label{zsphere}
Z_{\text{GR}}(S^{D})=(\mp i)^{D+2}|Z_{\text{GR}}(S^{D})| ,~~\text{with}~~~~~~~\frac{1}{\hbar}\rightarrow \frac{1}{\hbar}(1\pm i\epsilon)
\end{equation}

The fact that these path integrals have an overall phase seems to be in complete contradiction with their usual interpretation as partition functions \cite{Gibbons:1976ue}. However, since the gravitational path integral proved to be so reliable in other contexts, one might wonder if these phases have some physical significance. At least, we know their significance in some examples, such as the S-matrix in string theory \cite{Polchinski:1988ua, Polchinski:1998rq} and in $S^{p}\times M_{q}$ backgrounds where the phase knows about instabilities in the static patch of the $dS^{p}$ factor \cite{Ivo:2025yek}. 

Therefore, it seems reasonable to take the phase of the gravitational path integral seriously, in the hope that it has something to teach us about other observables in quantum gravity. Motivated by this, in this paper, we will study the Euclidean path integral over Coleman de Luccia instantons \cite{Coleman:1980aw} at the one-loop level, in particular clarifying its overall phase.

The relevance of studying these instantons is that their quantum field theory counterparts in flat space are famously related to false vacuum decay \cite{Coleman:1977py,Callan:1977pt}. More importantly, their tunneling interpretation is directly tied to the overall phase of the path integral over them, as stressed in \cite{Callan:1977pt}. To be more precise, the instantons responsible for vacuum decay in quantum field theory, called "bounces", have a single negative mode \cite{Callan:1977pt,Coleman:1987rm}. This implies that the path integral over them, $Z_{\phi}(\text{bounce})$, has an overall factor of $i$ \cite{Callan:1977pt}. Callan and Coleman \cite{Callan:1977pt} then argue that precisely because of this factor of $i$, these instantons give an imaginary energy shift to the false vacuum ground state, e.g, a decay width for the false vacuum.

The fact that the overall phase of these instantons is so directly tied to their physical interpretation makes them a natural laboratory for understanding the phase of the Euclidean gravitational path integral better. Namely, given that we know these instantons and their phase in quantum field theory, a natural question is what their counterpart is in the full gravity+matter path integral, more explicitly for Coleman de Luccia instantons \cite{Coleman:1980aw}. We refer to their contribution to the Euclidean path integral as $Z(\text{CdL})$. 

We should comment that the path integral in Coleman de Luccia instantons was studied in the past in many different papers \cite{Lavrelashvili:1985vn,Tanaka:1992zw,Garriga:1993fh,Tanaka:1998mp,Tanaka:1999pj,Lavrelashvili:1999sr,Khvedelidze:2000cp,Gratton:2000fj,Dunne:2006bt,Lee:2014uza,Koehn:2015hga}. However, it seems that a shared methodology in these papers is that, in evaluating the one-loop determinants of these path integrals, the authors seem to impose the gravitational constraints first. To be more specific, they study the path integral over "physical fluctuations" that satisfy the Hamiltonian and momentum constraints. 

A motivation for doing so is that imposing constraints first was often believed to be a solution to the conformal factor problem in at least pure gravity, which was recently disputed in \cite{Horowitz:2025zpx}. Another reason for imposing constraints first is that this is the natural thing to do if one sets up the path integral using the canonical formalism for gravity. Namely, taking the canonical formulation as a starting point and integrating first over the lapse and shift variables naturally imposes the gravitational constraints in the path integral.

By imposing the constraints first, one would expect the physical subspace of fluctuations of the Coleman de Luccia instantons to have a single negative mode, which should be the continuation of the single negative mode in quantum field theory. However, different choices of gauge fixing seem to give different answers as reviewed in \cite{Lee:2014uza}. Also, in some of the papers, such as in \cite{Lee:2014uza}, one seems to find more negative modes for these instantons than the single one we expect, even at small backreaction.

In this paper, we instead study the Euclidean path integral directly as an integral over metrics, which we gauge fix using the BRST procedure. To regularize the negative modes, we follow the prescription in \cite{Polchinski:1988ua,Maldacena:2024spf,Ivo:2025yek}. An advantage of working with the Euclidean path integral this way is that, while the number of negative modes of individual fields is gauge fixing dependent, the overall phase of the path integral seems to be gauge fixing independent. At least, this is the case for the pure gravity saddles studied in \cite{Ivo:2025yek,Shi:2025amq}. So, in this paper, we propose to use the overall phase of the path integral defined this way to figure out whether or not specific Coleman de Luccia solutions have a tunneling interpretation. 

In this paper, we are going to evaluate the Coleman de Luccia instanton contribution to the path integral, $Z(\text{CdL})$, at one-loop order, in the limit of small backreaction when the theory has a positive cosmological constant. We will show that it factorizes into a pure gravity sphere contribution $Z_{\text{GR}}(S^{D})$ and a quantum field theory contribution $Z_{\phi}(\text{bounce})$ as
\begin{equation}
\label{smbck}
Z(\text{CdL})=Z_{\text{GR}}(S^{D})Z_{\phi}(\text{bounce})(1+O(G_{N}))
\end{equation}

Most of the derivation is straightforward, but some zero modes of the background at zero backreaction become non-zero light modes of the Coleman de Luccia instanton. This is somehow reminiscent of the role of nearly zero modes in near extremal black holes mentioned in \cite{Banerjee:2023quv,Kapec:2023ruw,Blacker:2025zca}. Here, we focus on how these light-mode contributions reduce, at leading order, to what we would have obtained by treating them as exact zero modes.   

We will also propose a decay rate formula for the false vacuum in terms of gravitational path integrals. Such a formula for decay rate is, of course, well understood at the level of the action \cite{Coleman:1980aw}, but we propose how to fix its $O(1)$ coefficients as well. Using \nref{smbck} we will be able to verify that, as $G_{N}\rightarrow0$, the proposed formula for decay rates will reduce to a de Sitter analogue of the quantum field theory formula in \cite{Callan:1977pt}. This implies that, in this limit, if the quantum field theory path integral has the correct number of negative modes for tunneling, a small backreaction will not ruin it.

In the discussion section \nref{discu}, we also propose a prescription to determine how the phase changes as one makes the backreaction bigger.

In the rest of the introduction, we will summarize the main idea in more detail.

\subsection{A gravitational decay rate formula}
\label{decratesec}

\begin{figure}[h!]
    \centering
    \includegraphics[width=0.55\linewidth]{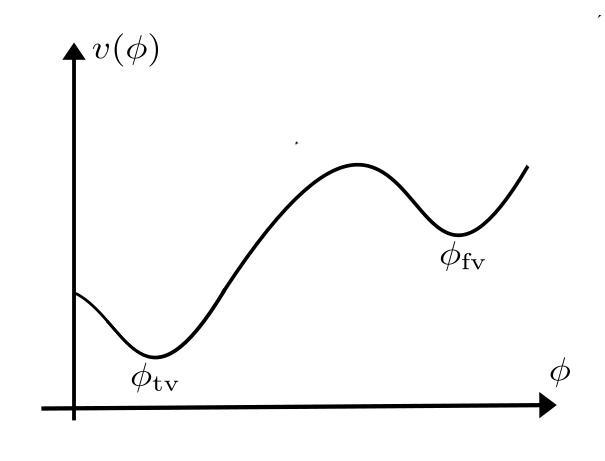}
    \caption{Scalar field potential $v(\phi)$ in the tunneling problem. $\phi_{\fv}$ is a false vacuum, and $\phi_{\tv}$ is the true vacuum}
    \label{scalarpotentialfig}
\end{figure}

Before further discussing a possible decay rate formula for the false vacuum that takes into account dynamical gravity, we first briefly review the quantum field theory formula by Callan and Coleman \cite{Callan:1977pt}. In their setup, they consider a scalar field theory in flat space, with a scalar potential $v(\phi)$. They assume the scalar potential has two local minima, $\phi_{\fv}$ and $\phi_{\tv}$. These correspond to the false and true vacuum, respectively, with $v(\phi_{\fv})>v(\phi_{\tv})$ by construction. In between $\phi_{\fv}$ and $\phi_{\tv}$, there is therefore a potential barrier in $v(\phi)$, as shown in Figure \nref{scalarpotentialfig}.

To understand the work done by Callan and Coleman, we need first to understand the "bounce" solutions. The bounce corresponds to a radially symmetric field configuration $\phi$ that is close to the true vacuum $\phi_{\tv}$ in the center of the solution, and asymptotes to the false vacuum $\phi_{\fv}$ away from it. These bounce solutions contribute as saddles to the Euclidean path integral. We refer to their contribution at the one-loop order as $Z_{\phi}(\text{bounce})$. One can furthermore argue that the one-loop determinant around these bounce saddles has an overall factor of $i$ \cite{Callan:1977pt}. By interpreting the bounce as a saddle contributing to an overlap of the Euclidean evolution operator $e^{-HT}$, \cite{Callan:1977pt} suggests that the overall factor of $i$ in $Z_{\phi}(\text{bounce})$ implies an imaginary energy shift to the false vacuum ground state energy $E_{\fv}$. In other words, the bounce saddles induce a decay width $\Gamma$ for the false vacuum. 

\begin{figure}[h!]
    \centering
    \includegraphics[width=0.9\linewidth]{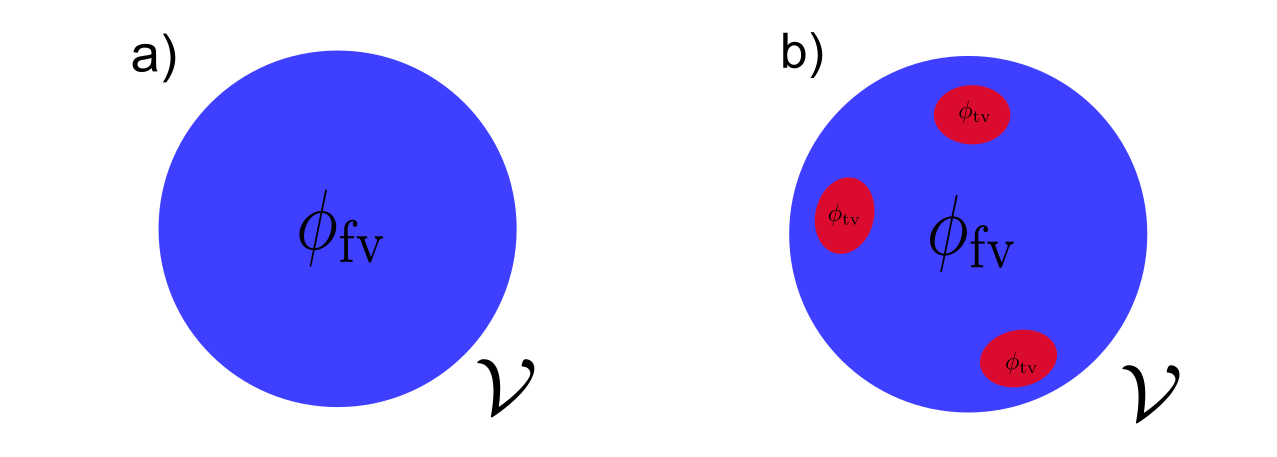}
    \caption{\textbf{a)} Region with spacetime volume $\mathcal{V}$ in a pure false vacuum configuration, painted in blue. \textbf{b)} Approximate classical configuration, built from the false vacuum configuration by adding bounces, in red, to some regions. The value of $\phi$ at the core of the red region is close to $\phi_{\tv}$, and it goes to $\phi_{\fv}$ at its boundary.}
\label{multibouncefig}
\end{figure}

The way that Callan and Coleman argue this is as follows: They take the theory to be defined in a subregion of flat space with large spacetime volume $\mathcal{V}$, and consider a saddle configuration where the scalar field is at the false vacuum, $\phi_{\fv}$, everywhere. Then, they compute the contribution of the bounces by summing over configurations built from the pure false vacuum one by filling multiple regions of $\mathcal{V}$ with the bounce field configuration, as shown in Figure \nref{multibouncefig}. They do this calculation in a "dilute gas" approximation, where they neglect interactions between the bounces. If the contribution of the pure false vacuum configuration for such a region is $Z_{\phi}(\phi_{\fv})$, the effect of summing over bounce insertions is
\begin{equation}
\label{zratioressum}
\sum_{n=0}^{\infty}\frac{1}{n!}\bigg(\frac{Z_{\phi}(\text{bounce})}{Z_{\phi}(\phi_{\fv})}\bigg)^{n}=e^{\frac{Z_{\phi}(\text{bounce})}{Z_{\phi}(\phi_{\fv})}}
\end{equation}

Using that $Z_{\phi}(\phi_{\fv})$ is real, one then only needs two features of $Z_{\phi}(\text{bounce})$ to argue that \nref{zratioressum} gives a decay rate for the false vacuum: First, $Z_{\phi}(\text{bounce})$ has a single negative mode, which gives it an overall factor of $i$. Secondly, the bounce has zero modes associated with translations of the instanton. This implies that there is a contribution to the one-loop determinant from integrating over the position of the center of the instanton, which gives an overall factor of $\mathcal{V}$ to $Z_{\phi}(\text{bounce})$. 

Therefore, one can rewrite $Z_{\phi}(\text{bounce})$ as $Z_{\phi}(\text{bounce})=i K\mathcal{V}Z_{\phi}(\phi_{\fv})$, with $K>0$ and independent of the spacetime volume $\mathcal{V}$. With this observation, from \nref{zratioressum} one can conclude that the sum over instantons contributes a factor $e^{i K \mathcal{V}}$ to the false vacuum partition function. To study forward Lorentzian time evolution, one should analytically continue the spacetime volume as $\mathcal{V}\rightarrow \mathcal{V}=i \mathcal{V}_{L}$, with $\mathcal{V}_{L}>0$. In that case, the effect of the sum over multiple instantons becomes an exponential decay $e^{-K\mathcal{V}_{L}}$. Since the exponential is proportional to the spacetime volume, the prefactor $K$ of the exponent is related to a decay rate per unit volume.  

This motivates Callan and Coleman to define a formula for the decay rate $\Gamma$ per unit volume of the false vacuum in terms of the single instanton contribution, $Z_{\phi}(\text{bounce})$, as\footnote{The perceptive reader will notice that formula \nref{gamqft} seems off by a factor of $2$ from the original definition in \cite{Callan:1977pt}. This is because we write the formula in terms of $Z_{\phi}(\text{bounce})$ defined in our paper, which is twice as big as the one defined in \cite{Callan:1977pt}. Therefore, the actual value for $\Gamma$ is the same. We explain the origin of the difference in more detail in section \nref{presc} and \nref{decft}.}
\begin{equation}
\label{gamqft}
\Gamma_{\text{QFT}}=\frac{1}{\mathcal{V}}\text{Im}\bigg(\frac{Z_{\phi}(\text{bounce})}{Z_{\phi}(\phi_{\fv})}\bigg)=K
\end{equation}

\begin{figure}[h!]
    \centering
    \includegraphics[width=0.8\linewidth]{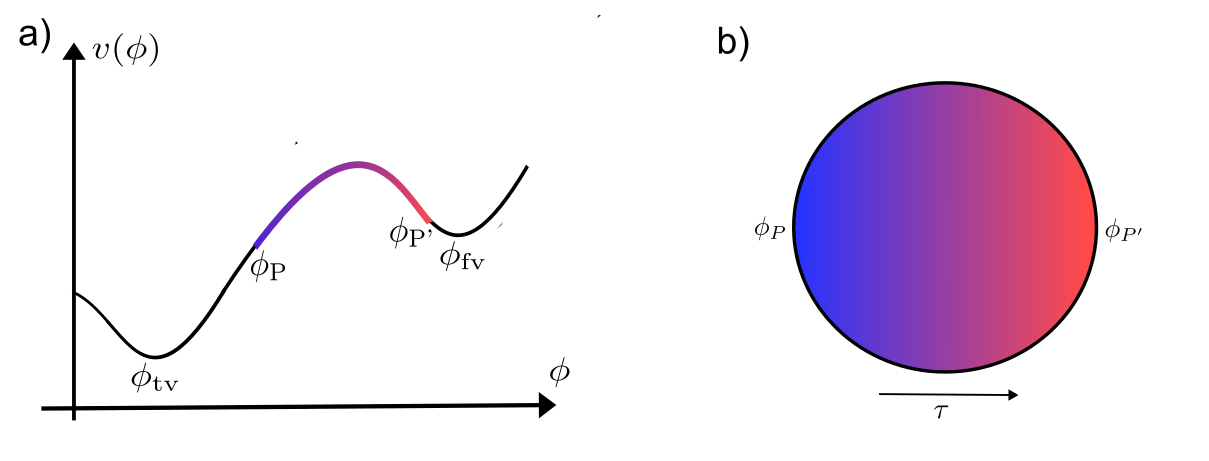}
    \caption{\textbf{a)}Plot of the scalar potential, with the scalar field configuration in the round sphere drawn in gradient. \textbf{b)} Visual display of the field configuration in the sphere. The configuration starts at a value $\phi_{\text{P}}$, in blue, at a pole $P$ of the sphere. Its value then changes along a radial variable $\tau$, until it arrives at the antipode $P'$ where the field has value $\phi_{\text{P'}}$, in red.}
    \label{BouncePlot}
\end{figure}

Also, note that the derivation leading to equation \nref{gamqft} is for a decay rate in flat space. However, we also assume that the formula holds in de Sitter, where the Euclidean background is the sphere, at least for a big enough de Sitter radius. In the round sphere $S^{D}$, the bounces are still radially symmetric solutions where $\phi$ crosses the potential barrier once. However, $\phi$ no longer goes all the way to $\phi_{\fv}$ since the sphere is a compact manifold. We display the form of these solutions in Figure \nref{BouncePlot}. Since de Sitter has a temperature, one can think of the decay rate defined by sphere path integrals as a thermally assisted vacuum decay \cite{Brown:2007sd}.

Having introduced and motivated the quantum field theory decay rate formula \nref{gamqft}, we can discuss the meaning of each term. First, equation \nref{gamqft} contains a ratio of partition functions, because the effect of each instanton in the path integral is to replace false vacuum regions with the nucleating bubbles. Secondly, equation \nref{gamqft} contains the imaginary part of these ratios because the ratio between $Z_{\phi}(\text{bounce})$ and $Z_{\phi}(\phi_{\fv})$ needs to be proportional to $i$ for the instanton sum to contribute as an exponential decay in Lorentzian time. The third, and last feature, of equation \nref{gamqft} is the division by the spacetime volume $\mathcal{V}$. The reasoning is that the decay happens everywhere in spacetime and should be interpreted as a uniform decay per unit time per unit volume. At a pragmatic level, this is also an important factor because it cancels against the factor of $\mathcal{V}$ in $Z_{\phi}(\text{bounce})$ and makes $\Gamma_{\text{QFT}}$ be independent of spacetime volume.

With all of these points in mind, we now make an educated guess of what the gravitational counterpart $\Gamma$ of \nref{gamqft} is. Because \nref{gamqft} is defined by the imaginary part of a ratio of path integrals, we expect that the same should be true for its gravitational counterpart. Therefore, we should replace the path integrals in \nref{gamqft} by their full gravity+matter counterparts. The configuration that generalizes the bounce once backreaction is taken into account is the Coleman de Luccia instanton \cite{Coleman:1980aw}. Its contribution to the path integral of gravity+matter to one-loop level is $Z(\text{CdL})$. For technical reasons, we also assume that the theory has a positive cosmological constant, and we study de Sitter to de Sitter transitions. The counterpart of $Z_{\phi}(\phi_{\fv})$ is therefore the gravity+matter path integral over a saddle where $\phi=\phi_{\fv}$ everywhere, and the geometry is a round sphere of appropriate radius. We call the path integral over this configuration $Z_{\fv}(S^{D})$.

From the third point, we should divide by some spacetime volume. This part of the generalization is unclear because we are no longer studying quantum field theory in a fixed spacetime. Therefore, the overall spacetime volume will depend on the specific field configuration. In particular, if we follow Callan and Coleman's idea of summing over multiple instanton configurations, the overall volume of the sphere will be sensitive to whether we have a single or numerous bounces.

For lack of a better choice, we divide by the spacetime volume of the round sphere geometry at the false vacuum\footnote{We would like to thank Juan Maldacena for discussions about this point.} $\phi_{\fv}$, since this is the analogue of the spacetime volume without any "bubble insertions". We call this volume $\text{Vol}(S_{\fv}^{D})$, with the prescription $\fv$ meaning the radius of the sphere takes into account the backreaction of the scalar potential $v(\phi_{\fv})$. With these considerations, we therefore propose the following formula for the decay rate of the false vacuum in terms of gravitational instantons
\begin{equation}
\label{gamint}
\Gamma=\frac{1}{\text{Vol}(S_{\fv}^{D})}\text{Im}\bigg(\frac{Z(\text{CdL})}{Z_{\fv}(S^{D})}\bigg)
\end{equation}

The term in equation\nref{gamint} coming from the ratio of partition functions is expected, at least at the level of the action contribution, from the principle of detailed balance \cite{Banks:2002nm}. Therefore, that contribution is expected on somewhat concrete grounds. The division by a volume in front is necessary for the correct zero backreaction limit, but whether or not $\text{Vol}(S_{\fv}^{D})$ is the proper choice for the volume is still unclear. Other volumes that match this one at zero backreaction would give the same limit.

Using formula \nref{smbck}, proved in the main discussion of the paper, we now argue that, if the theory has a fixed positive cosmological constant, as $G_{N} \rightarrow 0$ the $\Gamma$ defined for Coleman de Luccia instantons will reduce to a quantum field theory decay rate $\Gamma_{\text{QFT}}$ in de Sitter space. The reason is that the gravity+matter path integral for the false vacuum $Z_{\fv}(S^{D})$ also reduces to a product of path integrals in this limit. Namely, it reduces to the product of the pure gravity path integral in the round sphere $Z_{\text{GR}}(S^{D})$, and a scalar path integral around the pure false vacuum solution in the sphere, $Z_{\phi}(\phi_{\fv})$. With the choice of cosmological constant we will make, in the small backreaction limit, the volume of the round sphere at the false vacuum will reduce to the volume of the unit sphere, such that
\begin{equation}
\label{dechere}
\Gamma \approx \frac{1}{\text{Vol}(S^{D})}\text{Im}\bigg(\frac{Z_{GR}(S^{D})Z_{\phi}(\text{bounce})}{Z_{\text{GR}}(S^{D})Z_{\phi}(\phi_{\fv})}\bigg)=\frac{1}{\text{Vol}(S^{D})}\text{Im}\bigg(\frac{Z_{\phi}(\text{bounce})}{Z_{\phi}(\phi_{\fv})}\bigg)
\end{equation}

Therefore, in the small backreaction limit, the contribution from the gravitational path integrals $Z_{\text{GR}}(S^{D})$, and in particular their phases, cancels out. Thus, the decay rate formula reduces to that of a quantum field theory in de Sitter, and if $Z_{\phi}(\text{bounce})$ has a single negative mode, $\Gamma$ will be real and positive. 

One might perhaps have expected $Z(\text{CdL})$ to not have a tunneling interpretation, because of the extra phase in \nref{smbck} coming from $Z_{\text{GR}}(S^{D})$. The key, however, is that if we want to use $Z(\text{CdL})$ to define decay rates, we should also use similar Euclidean path integrals to define the false vacuum path integral in the denominator. Because the extra phase from gravity is also present in $Z_{\fv}(S^{D})$, the overall phase of their ratio is the usual $i$ coming from the quantum field theory bounce. It might seem puzzling that we are taking a ratio of complex partition functions to define the decay rate, but we take this to be a feature of Euclidean gravity, not a bug.

\subsection{Decay rate at small backreaction}

In this paper, we will consider in the setup the theory to consist of Einstein gravity with a minimally coupled scalar $\phi$, with a potential $v(\phi)$ as in section \nref{decratesec}. We also take the theory to have a fixed cosmological constant $\Lambda>0$. Since we are going to be interested in Coleman de Luccia instantons, both the background scalar and metric will be $SO(D)$ invariant. 

In particular, as one takes $G_{N}\rightarrow 0$, the metric will reduce to that of a round sphere $S^{D}$ with radius set by the positive cosmological constant $\Lambda$. For convenience, we take $\Lambda$ such that the sphere has a unit radius of curvature. Similarly, as $G_{N} \rightarrow 0$, the background scalar $\phi$ will limit to a bounce solution of the scalar equation of motion in the round sphere $S^{D}$. 

One might intuitively expect that we can simplify the calculation by treating the sphere $S^{D}$ as a fixed background, with the scalar field $\phi$ acting as a probe matter configuration in it. From this intuition, one might expect that the path integral contribution $Z(\text{CdL})$ decomposes into approximately the product of a pure gravity path integral in $S^{D}$, $Z_{\text{GR}}(S^{D})$, and a path integral contribution $Z_{\phi}(\text{bounce})$ from the scalar, obtained from treating it as a probe field in the fixed $S^{D}$ background. We could therefore wonder if the following formula is correct
\begin{equation}
\label{smbckq}
Z(\text{CdL}) \overset{?}{\approx} Z_{\text{GR}}(S^{D})Z_{\phi}(\text{bounce})
\end{equation}

We can argue that the intuitive formula \nref{smbckq} is correct more carefully by evaluating the full gravity+matter path integral to one-loop order in the small $G_{N}$ limit. At the level of the action contribution to the path integral, $e^{-I}$, it is clear that $Z(\text{CdL})$ reduces to $Z_{\text{GR}}(S^{D})Z_{\phi}(\text{bounce})$ as $G_{N} \rightarrow 0$. Therefore, the only thing left to argue is that the one-loop determinants also match at leading order in $G_{N}$.

We will generally write the one-loop determinants in terms of functional determinants, up to factors coming from zero modes of these operators that need to be taken into account separately. We expect most of these functional determinants to behave smoothly as $G_{N} \rightarrow 0$, so most of their contribution in the left-hand side of \nref{smbckq} should match the right-hand side. 

What can change, however, is that the backreaction might lift some zero modes. Because of that, some contributions treated as zero mode contributions in the right-hand side of equation \nref{smbckq} could be incorporated as coming from non-zero modes of functional determinants in the left-hand side of equation \nref{smbckq}. One might be confused, in particular, about what happens to some contributions to the path integral coming from isometries, such as the ones discussed in \cite{Anninos:2020hfj, Law:2020cpj}. To be more specific about what we mean: Since we are studying a theory with gravity, we have to do the path integral over fields and metrics in some gauge fixing scheme. 

If one works with a gauge fixing scheme that is based on local gauge conditions on the metric and fields, there is generally some residual gauge invariance. This is because if there is a coordinate transformation that leaves the classical solution invariant, it cannot be gauge fixed by a local condition on the metric and fields. Such a coordinate transformation is, in particular, an isometry. To take this unfixed gauge invariance into account, one has to divide manually by the path integral over this isometry group using the appropriate measure derived from the path integral, as stressed in \cite{Donnelly:2013tia, Anninos:2020hfj, Law:2020cpj}. 

Take the manifold to have an isometry group $G_{1}$ with dimension $|G_{1}|$ and canonical group volume $\text{Vol}(G_{1})_{c}$. Then, following the measure convention of \cite{Anninos:2020hfj, Law:2020cpj}, one can argue that the residual gauge group contribution to the path integral has the form
\begin{equation}
\label{ziso}
Z_{G_{1}}=\frac{1}{\text{Vol}(G_{1})_{c}}\big(AG_{N}\big)^{\frac{|G_{1}|}{2}} 
\end{equation}
with $A$ a constant fixed by the classical background. The confusion is then the following: The Coleman de Luccia instanton has a smaller isometry group, $SO(D)$, than the round sphere isometry group $SO(D+1)$. Therefore, the zero mode contribution from isometries in both sides of \nref{smbckq} is quite different. In particular, not even the factors of $G_{N}$ match. 

This implies that either the $G_{N}$ dependence of the path integral would change discontinuously once we take into account matter backreaction, and thus equation \nref{smbckq} would not hold, or there is another mechanism that gives rise to such factors $G_{N}$ and precisely cancels the discontinuity. We argue the latter to be true by finding the explicit mechanism. More explicitly, we argue that this discontinuity is taken into account by new non-zero modes of functional determinants in the left-hand side of equation \nref{smbckq}.

To understand the idea, we should note that the difference in isometry contributions between the left and right side of equation \nref{smbckq} comes from the fact that there are $D$ isometry generators in $SO(D+1)$ which are not in $SO(D)$. One can think of these generators as the $D$ coordinates in the sphere $S^{D}$, since $S^{D}$ is a coset of $SO(D+1)$ by $SO(D)$. In other words, the division by isometries in $Z_{\text{GR}}(S^{D})$ has, in comparison to the division by isometries in $Z(\text{CdL})$, an extra division by translations along the sphere.

However, $Z_{\phi}(\text{bounce})$ also has similar zero modes. This is because there are scalar zero modes associated with the fact that we can move the center of the instanton along the sphere. Since we integrate over translations in $Z_{\phi}(\text{bounce})$ and we divide by it in $Z_{\text{GR}}(S^{D})$, the volume integrals cancel out. Therefore, the overall contribution of these modes is a Jacobian factor from the ratio between the mode integrals in the numerator and denominator. This Jacobian factor scales with $G_{N}$ as $G_{N}^{\frac{D}{2}}$, because of the factors of $G_{N}$ in the associated isometry factors. Since there are $D$ independent translation generators, the Jacobian factor for each generator integral scales as $G_{N}^{\frac{1}{2}}$

We then only need to understand how $Z(\text{CdL})$ will reproduce these Jacobian factors. The point is that in the Coleman de Luccia instanton, the $D$ translation isometries of the round sphere will no longer be isometries, because they are broken, perturbatively in $G_{N}$, by the backreaction. So, they will instead be gauge fixed by an appropriate local gauge condition we choose. 

We choose the gauge condition to depend only on the metric at $G_{N}=0$, so that the metric and scalar path integrals decouple in the $G_{N} \rightarrow 0$ limit. Also, in the way that we do the gauge fixing in the paper, we add a gauge fixing term to the action. Therefore, the previously scalar zero modes are lifted by the gauge fixing term. The ghost zero modes associated with the broken isometries will also be lifted. Since these modes reduce to zero modes as $G_{N} \rightarrow 0$, they scale with $G_{N}$ and we call them "light" modes. 

These light modes appear in pairs because, in the gauge fixing scheme we adopt, they come from zero modes of both the gauge fixing ghosts and the matter degrees of freedom. More specifically, for the gauge fixing scheme we adopt in this paper the light eigenvalues of the ghost contribute to the ghost determinant as $|\lambda_{gh}| \sim G_{N}$ and the light eigenvalues of the scalar contribute to the path integral as $\lambda^{-\frac{1}{2}} \sim G_{N}^{-\frac{1}{2}}$. Moreover, the light scalar modes are positive definite, because, as we will show in the paper, their eigenvalue comes mainly from the gauge fixing term we add to the action, which is positive. Since these previously zero modes do not become negative, they will not affect our discussion of the phase. 

This implies that each light mode pair contributes to the path integral as an overall factor of $|\lambda_{gh}|\lambda^{-\frac{1}{2}} \sim O(G_{N}^{\frac{1}{2}})$. Therefore, they reproduce the Jacobian from the zero modes contribution in the right hand side of \nref{smbckq}, at the level of $G_{N}$ dependence. In the main discussion of the paper, we also argue more carefully that the $O(1)$ factors of the two sides of \nref{smbckq} match. More than that, we also argue that the corrections to the formula are $O(G_{N})$, such that
\begin{equation}
Z(\text{CdL})=Z_{\text{GR}}(S^{D})Z_{\phi}(\text{bounce})(1+O(G_{N}))
\end{equation}

The paper is structured as follows: In section \nref{presc}, we set up notation and review the basic machinery we use to compute functional determinants. In particular, we review how one should deal with negative modes and fields with an infinite number of them. We also highlight a difference between the analytic prescription for negative modes we introduce in that section and the original prescription of \cite{Callan:1977pt}. The difference between the prescriptions implies a relative factor of $2$ in the path integral over some negative modes. This point will be relevant in section \nref{decft}, and discussed again there.

In section \nref{decft}, we discuss the tunneling solutions in pure quantum field theory, and the one-loop determinant around them. This will help us motivate the definition of decay rate in the computation with dynamical gravity. In particular, we will see that requiring the correct sign for the decay rate $\Gamma$ implies that we need to analytically continue $\hbar$ in a specific way. 

In section \nref{vacdecgr}, we discuss the main results of the paper. We review the Coleman de Luccia instanton solutions, and we discuss the one-loop determinant around them. We also discuss how to compute the one-loop determinant at small backreaction by showing that equation \nref{smbck} holds. 

In section \nref{toyinstsec}, we introduce a simple instanton solution that we use to numerically verify the formula for the light mode eigenvalues derived in section \nref{vacdecgr}. The instanton introduced there is a nice toy model for Coleman de Luccia instantons, because for a special gauge fixing scheme, the fluctuations of the scalar field couple only to the trace part of the metric, so the one-loop determinant calculation is a bit simpler.

In section \nref{discu} we conclude the paper by discussing future directions. In particular, we propose a "Clutch prescription" that one can use to study how the phase of gravity+matter path integrals changes as we change the parameters in the theory.

\textbf{Note Added:} Since the first version of this paper, regulated computations of the phase in coupled gravitational systems have been further developed in \cite{Chen:2025jqm, Ivo:2025xek, Ivo:2026ijv}. Other papers studying the phase of the path integral more broadly were \cite{Giombi:2026sqa, Shi:2026wnk, Ookouchi:2026bqj}. In particular, \cite{Shi:2026wnk} studied the phase of Hawking-Moss instantons, closely associated with the instantons we discuss here. \cite{Ookouchi:2026bqj} subsequently studied the phase of Coleman-de Luccia instantons numerically at finite backreaction using a different approach.

\section{Preliminary discussion}
\label{presc}

In this section, we discuss the general formalism we use to compute one-loop determinants throughout the paper. The main motivation is to clarify notation and to discuss the prescription for how to deal with negative modes as in \cite{Polchinski:1988ua,Ivo:2025yek,Shi:2025amq}.

\subsection{One-loop determinants}

Throughout the paper, when we discuss computing path integrals, we will generally mean that we are doing so via saddle point approximation. We will generally be interested in the contribution of a specific saddle, and we are going to evaluate its contribution at the one-loop level by discussing both its action and the one-loop determinant around the solution. We will denote the contribution of a specific saddle by the symbol $Z$ with appropriate extra clarifiers.

Let us denote the fields running in the path integral collectively by $\Phi$, the action of a saddle by $I$, and the quadratic action around it by $\delta_{2}I$. Then, the contribution $Z$ of that saddle at one-loop level reads
\begin{equation}
Z=e^{-I}\int d\Phi\,e^{-\delta_{2}I}=e^{-I}Z_{1-\text{loop}}
\end{equation}
where we call the one-loop determinant contribution $Z_{1-\text{loop}}$. To compute the one-loop determinant, one therefore needs to find the quadratic action $\delta_{2}I$. Taking the background geometry where the classical solution is defined to be a $D$-dimensional manifold $\mathcal{M}$, we will generally denote integrals over the manifold by $\int$ without further specification, but it will always implicitly mean
\begin{equation}
\int =\int_{\mathcal{M}}d^{D}x\,\sqrt{g}
\end{equation}

The quadratic action $\delta_{2}I$ can be written generally in terms of a local integral over the fluctuation fields. After writing $\delta_{2}I$ in terms of such an integral, we can furthermore write it as an overlap of some fluctuation operator $M$, of the form
\begin{equation}
\label{quadac}
\delta_{2}I=\frac{1}{2}(\Phi,M\Phi)
\end{equation}
with $(\bullet,\bullet)$ some appropriate local norm. We always define the local norm, for a field with spacetime indices, for example $\phi_{a_{1}....a_{n}}$, via
\begin{equation}
\label{locnormdi}
(\phi_{a_{1}....a_{n}},\phi_{a_{1}....a_{n}})=\int \phi_{a_{1}...a_{n}}\phi_{b_{1}...b_{n}}g^{a_{1}b_{1}}...g^{a_{n}b_{n}}
\end{equation}

That is, we contract the field $\phi$ with itself using the background metric, and we integrate it over the manifold. The definition of the norm for fields with upper indices, or with both upper and lower indices, is the obvious counterpart of \nref{locnormdi}.

The last step to do the functional integration is to define a measure $D\Phi$ for the path integral. We will use the measure following from the local norm \nref{locnormdi} with appropriate index contractions, and normalized such that
\begin{equation}
\label{measgn}
\int D\Phi\,e^{-\frac{1}{2}(\Phi,\Phi)}=1
\end{equation}
where $\Phi$ could be a field with indices, but we omitted them for simplicity.

To write the measure more explicitly, it is useful to expand $\phi$ in a local basis of functions $f_{n}$ of the manifold, normalized such that $(f_{n},f_{m})=\delta_{nm}$, that is
\begin{equation}
\Phi=\sum_{n} \Phi_{n}f_{n} \rightarrow (\Phi,\Phi)=\sum_{n}\Phi_{n}^{2}
\end{equation}

We can then derive the measure compatible with \nref{measgn} to be
\begin{equation}
\label{measgnc}
D\Phi=\prod_{n} \frac{d\Phi_{n}}{\sqrt{2\pi}}
\end{equation}

We should note that while we omitted the indices in $\Phi$, the basis functions $f_{n}$ in the expansion will, of course, depend on their precise form. Note that measures following from the local norm but differing from \nref{measgn} only in the normalization condition will differ from \nref{measgnc} by a constant in each $c_{n}$ integral. Therefore, the difference is an infinite constant, but it is a local contribution that can be absorbed into appropriate counterterms and should not change the physics. Therefore, the specific normalization \nref{measgn} is not necessary, just convenient.

We can pick the basis $f_{n}$ we expand $\Phi$ over to be the one that diagonalizes the operator $M$ in \nref{quadac}. Then, assuming $M$ has a set of eigenvalues $\lambda_{n}$, one can evaluate the path integral by doing the appropriate Gaussian integrals. However, if $M$ has zero modes, we cannot do the Gaussian integral over them. The integral over zero modes, then, is a bit different and needs to be treated separately. We call it $Z_{0}$. Having this result, we can evaluate the path integral as
\begin{equation}
Z=e^{-I}\int D\phi\,e^{-\frac{1}{2}(\phi,M\phi)}=Z_{0}\bigg(\prod_{n}'\lambda_{n}^{-\frac{1}{2}}\bigg)e^{-I}=Z_{0}Z'
\end{equation}
with $\prod'$ meaning the product over eigenvalues with zero modes omitted, and where we defined 
\begin{equation}
\label{zprime}
Z'=\bigg(\prod_{n}'\lambda_{n}^{-\frac{1}{2}}\bigg)e^{-I}
\end{equation}
to be the contribution to the path integral coming from non-zero modes and the on-shell action. We will use the notation $Z'$ to mean that for general path integrals. Note that while the path integral over the zero modes $Z_{0}$ seems to be divergent at the naive level, when the zero modes are associated with a compact symmetry group, they will have a compact range and the integral over them will be finite.

Another point to worry about is that if $M$ has negative modes, the Gaussian integral over them will not converge, so we need to regularize them somehow. For that, it is perhaps instructive to consider a single Gaussian integral first, such as in \cite{Maldacena:2024spf,Ivo:2025yek}, of the form
\begin{equation}
\int_{-\infty}^{\infty} dx\, e^{(1\pm i\epsilon )x^{2}}
\end{equation}

If $\epsilon$ were zero, the integral above would be ill-defined, but for $\epsilon$ non-zero, we can define the integral by analytic continuation. That is, we can make the integral convergent by rotating the contour of $x$ while avoiding the direction of maximum increase. 

Another equivalent perspective is that we can compute the integral for when the coefficient of the Gaussian is convergent, and we then analytically continue the coefficients of the Gaussian. In particular
\begin{equation}
\int_{-\infty}^{\infty}dx\,e^{(1\pm i\epsilon)x^{2}}=(\pm i)\sqrt{\pi}
\end{equation}

So once we fixed an $i\epsilon$ prescription, this Gaussian integral, defined by analytic continuation, is well defined and finite. We can implement this prescription naturally when computing one-loop determinants by analytically continuing the Planck constant $\hbar$ of the theory to be slightly complex as
\begin{equation}
\label{iepsilonp}
\frac{1}{\hbar} \rightarrow \frac{1}{\hbar}(1\pm i \epsilon)
\end{equation}

In quantum field theory, each field mode in \nref{measgnc} with a negative eigenvalue of $M$ will be a wrong-sign Gaussian. If $M$ has a finite number of negative modes $n_{E}$, the integral over them is regularized via the prescription described above, and we obtain an overall phase in the path integral of 
\begin{equation}
\label{negphase}
Z=(\pm i)^{n_{E}}|Z|
\end{equation}

Note that this prescription is slightly different from the negative modes prescription introduced in \cite{Callan:1977pt}. To be more specific, the bounce in \cite{Callan:1977pt} has a single negative mode, which contributes an extra factor of $\frac{1}{2}$ on top of the factor of $(\pm i)$ we discussed. The way this factor of $\frac{1}{2}$ appears is because they define the path integral for unstable saddles by analytic continuation of the potential. This would imply that we should pick the contour of integration through the unstable saddle in a specific way, which would give the factor of $\frac{1}{2}$. 

Therefore, the prescription for negative modes in \cite{Maldacena:2024spf, Ivo:2025yek}, which we reviewed in this section, seems to be incomplete. If this factor of $\frac{1}{2}$ is supposed to be included for some of the negative modes of the saddle in question, then their contribution in equation \nref{negphase} would have to be corrected to be $(\frac{\pm i}{2})^{n_{E}}$.

\subsection{Fields with an infinite number of negative modes}
\label{inftneg}

While \nref{iepsilonp} regularizes the path integral in the presence of a finite number of negative modes, there are additional subtleties when there are infinitely many such modes. Naively, it seems one would obtain an infinite phase of $(\pm i)^{\infty}$ from applying the regularization. The idea, however, is that if one rotated an entire field $\phi(x)$ as
\begin{equation}
\phi'(x)=i \phi(x)
\end{equation}
it would be just an ultralocal field redefinition \cite{Polchinski:1988ua,Hawking:2010nzr}, which can be absorbed into local counterterms of the action. Note this is the same as rotating all the basis modes of $\phi(x)$ as $\phi_{n} \rightarrow i \phi_{n}$. 

Therefore, if all the modes in the basis expansion of $\phi$ were negative, we could absorb the full phase in local counterterms, leaving a zero finite part for the phase.  This also implies that if the field has an infinite number of negative modes, but a finite number $n_{m}$ of the modes are non-negative, then the phase has an overall finite part. 

To see this, at a heuristic level let us refer to the product over one factor of $(\pm i)$ per mode in the expansion as $(\pm  i)^{n_{all}}$, that is $\prod_{n} (\pm i)=(\pm i)^{n_{all}}$. If a field has all modes in its expansion negative, except for $n_{m}$ modes which are non-negative, then its overall phase is
\begin{equation}
\prod_{\text{neg}}(\pm i)=(\mp i)^{n_{m}}\prod_{n}(\pm i)=(\mp i)^{n_{m}}(\pm i)^{n_{\text{all}}}
\end{equation}
where we used the product $\prod_{\text{neg}}$ to mean the product is only over the negative modes. Then, we used that we can think of this phase as one $\pm i$ per mode in the expansion, that can be absorbed into local counterterms, minus a finite subtraction for the modes missing from the product. This implies that a mostly negative field with $n_{m}$ non-negative modes leads to an overall finite phase of
\begin{equation}
\label{minusneg}
(\mp i)^{n_{m}}
\end{equation}

Note that the contribution per missing mode is the opposite of the contribution of a negative mode for a mostly positive field. This is important for gauge fixing invariance of the phase as discussed in \cite{Ivo:2025yek,Shi:2025amq}.

Another subtlety is that if we take the number of modes in an eigenbasis to be literally infinite, then the number of "missing modes" can be ambiguous. That is, there is an isomorphism from the basis without missing modes to the one with missing modes. 

To make the above discussion more well-defined, one can imagine instead that we regularize the theory in such a way that the number of basis modes of a field is big but finite. In this case, the concept of missing modes would be well defined. It would be interesting to study such a regularization more carefully and understand how/if the phase in equation \nref{minusneg} emerges.

Another subtlety is whether the factors of $\frac{1}{2}$ from the prescription in \cite{Callan:1977pt} should be included here. That is, if we believed that these factors of $\frac{1}{2}$ should be present for all negative mode contributions of a specific mostly negative field, their contribution in equation \nref{minusneg} would need to be corrected. Since each negative mode would come with an extra $\frac{1}{2}$, missing negative modes from that mostly negative field would come with a factor of $2$. Therefore, their contribution in equation \nref{minusneg} would be corrected to $(\mp 2i)^{n_{m}}$.

\section{Vacuum decay in quantum field theory}
\label{decft}

Here we discuss tunneling saddles in quantum field theory. We start in section \nref{flatsol} by reviewing the tunneling solutions in flat space \cite{Coleman:1977py,Callan:1977pt}, where their decay rate interpretation is clearer. This will help us motivate the definition of decay rate in terms of Euclidean path integrals in more general situations where its interpretation is not as clear. 

We then also discuss the sphere counterparts of these solutions, which are nothing but the zero backreaction limit of Coleman de Luccia instantons introduced in \cite{Coleman:1980aw}. This is going to give us the intuition necessary to study the instanton with dynamical gravity. 

The theory we will be discussing is of a canonically normalized scalar field with some potential $v(\phi)$. The Euclidean action is
\begin{equation}
I_{\phi}=\int \bigg(\frac{1}{2}(\n \phi)^{2}+v(\phi)\bigg)
\end{equation}
and the classical solutions will respect the equation of motion
\begin{equation}
-\n^{2}\phi+v'(\phi)=0
\end{equation}
with $'=\partial_{\phi}$ and $\n^{2}$ the Laplacian of the appropriate background geometry.

We also make some assumptions about the potential. Namely, we assume that the potential has a local maximum as in Figure \nref{scalarpotentialfig}, e.g., a potential barrier, separating two different vacua. We call these vacua $\phi_{\fv}$, to stand for the false vacuum, and $\phi_{\tv}$, to stand for the true vacuum. The common feature of the tunneling solutions we study in both flat space and the sphere is that the field $\phi$ will start from the side of the potential barrier where the true vacuum $\phi_{\tv}$ is, and will flow monotonically to the side of the barrier where the false vacuum $\phi_{\fv}$ is, crossing the top of the potential barrier only once.

\subsection{Flat space solutions}
\label{flatsol}

Here we review the bubble nucleation saddles in flat space introduced in \cite{Coleman:1977py}, and their one-loop determinant discussed in \cite{Callan:1977pt}. The line element in FRW form is
\begin{equation}
ds^{2}=d\tau^{2}+\tau^{2}d\Omega_{D-1}^{2}
\end{equation}
with $d\Omega_{D-1}^{2}$ the line element in $S^{D-1}$.

The classical field configuration $\phi$ in these solutions is only dependent on the radial variable $\tau$, and it goes monotonically from the true vacuum side of the barrier at $\tau=0$ to $\phi=\phi_{\fv}$ at $\tau=\infty$. To be more precise, the boundary condition at $\tau=0$ is $\dot{\phi}(0)=0$. The value of $\phi$ far enough from $\tau=0$ is very close to $\phi_{\fv}$, since it asymptotes to it at infinity. One therefore expects that the non-trivial part of the field configuration should be localized in a finite region of spacetime. 

In particular, we expect the difference between the action $I_{\phi}(\text{bounce})$ of this solution and the action $I_{\phi}(\phi_{\fv})$ of a solution with $\phi=\phi_{\text{fv}}$ everywhere to be finite. We call this action difference $B$, more explicitly
\begin{equation}
B=I_{\phi}(\text{bounce})-I_{\phi}(\phi_{\text{fv}})=\int \bigg(\frac{1}{2}(\n \phi)^{2}+v(\phi)-v(\phi_{\text{fv}})\bigg)
\end{equation}

The next relevant question is what the one-loop determinant around the bounce solutions is. The quadratic action around this saddle is given by
\begin{equation}
\label{qdacphi}
\delta_{2}I_{\phi}=\int \frac{1}{2}\varphi(-\n^{2}+v''(\phi))\varphi=\frac{1}{2}(\varphi,M_{\phi}\varphi)
\end{equation}
where we defined the fluctuation operator $M_{\phi}$
\begin{equation}
M_{\phi}=-\n^{2}+v''(\phi)
\end{equation}
and the local norm is
\begin{equation}
\label{sclocnorm}
(\varphi,\varphi)=\int \varphi^{2}
\end{equation}

One can show that the operator $M_{\phi}$ has a single negative mode \cite{Coleman:1987rm}, and we assume its only zero modes are the ones that follow from symmetry. There are $D$ zero modes that follow from symmetry, one for each isometry of the background metric that changes the background field $\phi$ non-trivially. More specifically, they correspond to translations of the center of the instanton. 

To compute the one-loop determinant, we have to define a local measure for the path integral. We take the measure derived from the local norm \nref{sclocnorm}, normalized such that
\begin{equation}
1=\int D\varphi\,e^{-\frac{1}{2}(\varphi,\varphi)}
\end{equation}

Expanding the fields $\varphi$ into a complete basis $f_{n}$ of eigenmodes of $M_{\phi}$ as $\varphi=\sum_{n}\phi_{n}f_{n}$, we have that
\begin{equation}
D\varphi=\prod_{n} \frac{d\phi_{n}}{\sqrt{2\pi}}
\end{equation}
where we took $f_{n}$ such that $(f_{n},f_{m})=\delta_{nm}$. If there were no zero modes, because of our choice of normalization, the one-loop determinant would be given by a product over eigenvalues of the fluctuation operator $M_{\phi}$. Namely if $M_{\phi}$ had a set of eigenvalues $\lambda_{n}$, the one-loop determinant would be $\prod_{n} \lambda_{n}^{-\frac{1}{2}}$. However, $M_{\phi}$ has some zero modes that we have to treat differently, and the result for the path integral, which we call $Z_{\phi}(\text{bounce})$, is
\begin{equation}
\label{pre1loopqft}
Z_{\phi}(\text{bounce})=e^{-I_{\phi}(\text{bounce})}\int D\varphi\,e^{-\frac{1}{2}(\varphi,M_{\phi}\varphi)}=Z_{\phi,0}\text{Det'}(-\n^{2}+v'')^{-\frac{1}{2}}e^{-I_{\phi}(\text{bounce})}
\end{equation}

We define $Z_{\phi,0}$ to be the integral over zero modes and $\text{Det'}$ refers to the functional determinant of an operator with zero modes omitted, or equivalently the product over its non-zero eigenvalues
\begin{equation}
\text{Det}'M_{\phi}=\prod_{n}'\, \lambda_{n}
\end{equation}
with $\prod_{n}'$ meaning the product over the modes with zero modes excluded.

The zero modes of $M_{\phi}$ have the form
\begin{equation}
\varphi=\xi^{a}\n_{a}\phi
\end{equation}
with $\xi^{a}$ an isometry of the background that corresponds to translations. In flat space, they can be very clearly written as translations along the Cartesian coordinates. More explicitly, for $\xi^{a}=(\Delta x)\partial_{x}$ the fluctuation $\varphi$ corresponds to shifting the instanton along the $x$ direction by a constant $\Delta x$. 

We can therefore relate the path integral over the zero modes to an integral over the center of the instanton. To do so, we relate the mode corresponding to shifting the instanton by $\Delta x$ to the unit normalized modes $\phi_{0}$ by writing
\begin{equation}
\phi_{0}^{2}=\int \varphi^{2}=(\Delta x)^{2}\int \bigg(\frac{d\tau}{dx}\bigg)^{2} \bigg(\frac{d\phi}{d\tau}\bigg)^{2}=(\Delta x)^{2}\bigg(\frac{1}{D}\int \dot{\phi}^{2}(\tau)\bigg) \rightarrow d\phi_{0}=\bigg(\frac{1}{D}\int \dot{\phi}^{2}\bigg)^{\frac{1}{2}}dx
\end{equation}
where we did the integral over the angles to replace $\bigg(\frac{d\tau}{dx}\bigg)^{2}$ by its averaged value over a given $S^{D-1}$. The integral over the zero modes is then related in a straightforward way to the integral over the position of the center of the instanton as
\begin{equation}
\label{flatsol0}
Z_{\phi,0}=\bigg(\frac{1}{2\pi D}\int \dot{\phi}^{2}\bigg)^{\frac{D}{2}}\int d^{D}x
\end{equation}

Note that in solving for the instanton, we assumed the solution was in infinite volume, which implies the integral over the translations is divergent. One can regularize them by taking the background to have a finite but very large spacetime volume $VT$. Since the non-trivial part of the instanton is localized in a finite volume, we can approximate the regularized volume solutions by the infinite volume ones we discussed. At this level of approximation, the integral over zero modes follows as before with a fixed regularized spacetime volume $VT$.

We also have to deal with the negative eigenvalue in $M_{\phi}$. According to the prescription we reviewed in section \nref{presc}, we would rotate the mode and obtain an $\pm i$ depending on how we continued $\hbar$, and obtain
\begin{equation}
\label{1loopqft}
Z_{\phi}(\text{bounce})=(\pm i) VT\bigg(\frac{1}{2\pi D}\int \dot{\phi}^{2}\bigg)^{\frac{D}{2}}\, \text{Det}'|(-\n^{2}+v''(\phi))|^{-\frac{1}{2}}e^{-B-I_{\phi}(\phi_{\text{fv}})}
\end{equation}
where $\text{Det}'$ is the functional determinant with zero modes omitted, and we took the absolute value of the fluctuation operator since we dealt with the factors of $i$ separately. 

The result in \nref{1loopqft} is almost the same as in \cite{Callan:1977pt}. According to \cite{Callan:1977pt}, the actual contribution of the bounce at one-loop level is $\frac{1}{2}$ of \nref{1loopqft}, because of their prescription for dealing with the negative mode. However, we proceed using \nref{1loopqft}, which followed from the prescription of section \nref{presc}, and we postpone the discussion of this difference to the end of this subsection, where we will correct these factors appropriately.

To understand the relation of this path integral to decay rates, note that the contribution we discussed in \nref{1loopqft} is for a single instanton. Still, in a large spacetime volume, we can have multiple instantons. In the limit of small instanton density, we can ignore their interaction and treat them as independent contributions. So, to take into account multi-instanton configurations, we sum over configurations that are false vacuum almost everywhere except for a few finite regions where we have the instantons.

In replacing a region of this fixed spacetime with the instanton, we are omitting the contribution to the path integral we would have obtained from a pure false vacuum configuration there, which we call $Z_{\phi}(\phi_{\fv})$. So, in summing over multiple instantons, we should think of the contribution of inserting one instanton as a ratio between $Z_{\phi}(\text{bounce})$ and $Z_{\phi}(\phi_{\fv})$, which has the overall form
\begin{equation}
\frac{Z_{\phi}({\text{bounce})}}{Z_{\phi}(\phi_{\text{fv}})}=(\pm i) VT\bigg(\frac{1}{2\pi D}\int \dot{\phi}^{2}\bigg)^{\frac{D}{2}}\, \frac{\text{Det}'|(-\n^{2}+v''(\phi))|^{-\frac{1}{2}}}{\text{Det}(-\n^{2}+v''(\phi_{\text{fv}}))^{-\frac{1}{2}}}e^{-B}=(\pm i)VTK
\end{equation}
where we defined
\begin{equation}
K=\bigg(\frac{1}{2\pi D}\int \dot{\phi}^{2}\bigg)^{\frac{D}{2}}\, \frac{\text{Det}'|(-\n^{2}+v''(\phi))|^{-\frac{1}{2}}}{\text{Det}(-\n^{2}+v''(\phi_{\text{fv}}))^{-\frac{1}{2}}}e^{-B}
\end{equation}

Note that the functional determinant in the denominator came from the fluctuation operator around the pure false vacuum solution. The contribution from $n$ instantons needs to be divided by a factor of $n!$ since they are identical, which allows us to resum the multi-instanton contributions as
\begin{equation}
\label{expdec}
\sum_{n=0}^{\infty}\frac{1}{n!}\bigg((\pm i)VTK\bigg)^{n}=e^{(\pm i)VTK}
\end{equation}

Note that this sum would be an overall phase for a real spacetime volume $VT$. If we analytically continue the saddles to positive Lorentzian time, however, we take $T \rightarrow iT$ and we obtain that the overall result is an exponentially decaying contribution instead. In doing that, we also needed to pick a specific sign for the $(\pm i)$ contribution, namely the $+i$ sign, such that the contribution is exponentially decaying instead of exponentially growing. In other words, in the language of section \nref{presc}, requiring exponential decay forces us to analytically continue $\hbar$ as
\begin{equation}
\frac{1}{\hbar}\rightarrow\frac{1}{\hbar}(1+i\epsilon)
\end{equation}
which is the opposite convention chosen in \cite{Maldacena:2024spf,Ivo:2025yek}. 

It is then clear that the contribution from the instantons in \nref{expdec} makes the false vacuum saddle exponentially decaying in Lorentzian time $T$. Here, following \cite{Callan:1977pt}, one would then try to interpret \nref{expdec} as the contribution of the false vacuum ground state to an overlap function of $e^{-iHT}$. Then, the exponentially decaying piece would be a decay width term $e^{-\frac{\Gamma}{2} VT}$, with $\Gamma$ being a decay rate per unit volume. The $\frac{1}{2}$ in $\Gamma$ follows because such an overlap is an amplitude, and the probability goes as the square of that. 

If we tried to interpret the "naive" \nref{expdec}, computed with our $Z_{\phi}(\text{bounce})$ from \nref{1loopqft}, as an overlap of this form, we would obtain a decay rate $\Gamma$ twice as big as it should be according to \cite{Callan:1977pt}. The reason is that the $Z_{\text{bounce}}$ we found in \nref{1loopqft} does not include a subtle factor of $\frac{1}{2}$ discussed in \cite{Callan:1977pt}. To understand this factor of $\frac{1}{2}$ in \cite{Callan:1977pt}, let us briefly discuss their prescription for dealing with the negative mode in the bounce.

\begin{figure}[h!]
    \centering
    \includegraphics[width=0.8\linewidth]{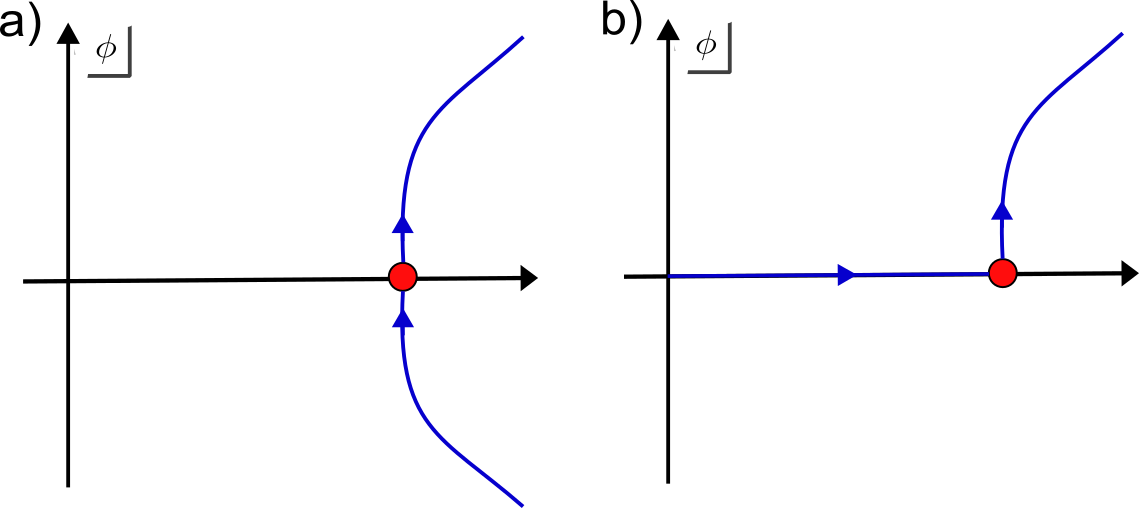}
    \caption{Two possible steepest descent contours going through a "bubble saddle", which we denoted by a red point. \textbf{a)} This is the "naive" contour through the bubble. Using the prescription of section \nref{presc}, we would do the integral over the entire blue contour and obtain the negative mode contribution discussed there. \textbf{b)} This is a rough drawing of the contour through the bubble discussed in \cite{Callan:1977pt}. At the red point, where the contour goes through the bubble saddle, one should imagine the curve is actually smoothed out a little bit. The negative mode contribution to the one-loop determinant comes only from the vertical line emanating from the red point; therefore, it is only half of the contribution from the "naive" contour. This is explained in much more detail in the nice work of \cite{Andreassen:2016cvx}.}
    \label{contours}
\end{figure}

Callan and Coleman's prescription is motivated by how one would carefully define the bounce contribution through analytic continuation of the potential. That is, they imagine that the path integral can be done in some region of parameter space where all the modes around the saddle are positive. Then, they analytically continue the parameters of the potential until they get the path integral over the bounce saddle. By doing this procedure carefully, one can justify that the negative mode contour going through the bounce saddle has the overall form displayed in the second contour of Figure \nref{contours}. Because of the form of the contour, the one-loop determinant is $\frac{1}{2}$ of what one might naively have expected.

Their prescription for dealing with the negative mode in the bounce is the correct one, as one can check by studying vacuum decay from different perspectives, as in \cite{Andreassen:2016cvx}. Unfortunately, our formalism discussed in section \nref{presc} does not seem to be able to reproduce this factor of $\frac{1}{2}$, and because of that our $Z_{\phi}(\text{bounce})$ was twice as big as it should have been. Therefore, the prescription for negative modes in section \nref{presc} is somehow incomplete for these instantons. To obtain correct decay rates, we then need to introduce a factor of $\frac{1}{2}$ by hand in front of $Z_{\phi}(\text{bounce})$. Another possibility is that in using the prescription for negative modes in section \nref{presc} we are not computing the same physical quantity, namely the overlap of $e^{-HT}$, that was discussed in \cite{Callan:1977pt}.

We remain agnostic about what the correct interpretation of the difference is, but at a pragmatic level: The correct definition of decay rate, that matches with \cite{Callan:1977pt}, and is written in terms of the "naive" Euclidean path integrals evaluated using our prescription, such as $Z_{\phi}(\text{bounce})$, is
\begin{equation}
\label{decrate}
\Gamma_{\text{QFT}} =\frac{2}{VT}\text{Im}\bigg(\frac{\frac{1}{2}Z_{\phi}(\text{bounce})}{Z_{\phi}(\phi_{\text{fv}})}\bigg)=\frac{1}{VT}\text{Im}\bigg(\frac{Z_{\phi}(\text{bounce})}{Z_{\phi}(\phi_{\text{fv}})}\bigg)=K
\end{equation}
where the ratio is between the single bubble contribution to the path integral we computed in \nref{1loopqft}, and the pure false vacuum contribution. 

Note that the $2$ in the first line of \nref{decrate} comes because the decay width contribution is $e^{-\frac{\Gamma}{2}VT}$. The factor of $\frac{1}{2}$ in front of $Z_{\phi}(\text{bounce})$, however, was introduced by hand to match the subtle factor of $\frac{1}{2}$ discussed in \cite{Callan:1977pt}. These factors of $2$ do not seem to have anything to do with each other, but they cancel out.

Later, we also use the definition \nref{decrate} as inspiration to define the decay rate in cases where gravity is dynamical. In these cases, it seems that the natural extension of \nref{decrate} is to replace the spacetime volume $VT$ by the spacetime volume of the relevant false vacuum geometry, and to replace the scalar path integrals $Z_{\phi}$ in \nref{decrate} by their full gravity+matter path integral counterparts. 

These gravitational path integrals will all still be evaluated using the prescription for negative modes in section \nref{presc}. But again, we will define their decay rates as something like \nref{decrate} so that we can avoid including the subtle factors of $\frac{1}{2}$ in the definition of the path integrals $Z$.

\subsection{Sphere solutions}
\label{nobcksphere}

Another interesting background geometry for these instantons is a sphere, which will be the relevant background for our discussion with dynamical gravity in section \nref{vacdecgr}. We take the line element to be
\begin{equation}
ds^{2}=d\tau^{2}+\sin^{2}\tau\,d\Omega_{D-1}^{2}
\end{equation}
where, to avoid clutter, we set the radius of the sphere to one, but it is simple to restore it whenever necessary. The instanton solution we consider is still a $SO(D)$ symmetric field configuration $\phi(\tau)$ that interpolates monotonically from one side of a potential barrier to the opposite side. It satisfies the equation of motion
\begin{equation}
-\n^{2}\phi+\partial_{\phi}v(\phi)=-\ddot{\phi}-(D-1)\cot \tau\, \dot{\phi}+\partial_{\phi}v(\phi)=0
\end{equation}

Since the space is compact, the field configuration cannot go from one vacuum to the other exactly. However, the necessary boundary condition is only regularity of the solution, which, for example, implies that $\dot{\phi}=0$ at $\tau=0$ and $\tau=\pi$. 

The rest of the discussion goes the same as in flat space, but $B$ is now the difference between the bounce action and the action of the pure false vacuum configuration on the sphere, namely
\begin{equation}
B=I_{\text{bounce}}-I(\phi_{\fv})=\int \bigg(\frac{1}{2}(\n \phi)^{2}+v(\phi)-v(\phi_{\fv})\bigg)
\end{equation}

Another difference from flat space is that in the sphere, a regular monotonic bounce need not exist for an arbitrary potential \cite{Hawking:1981fz,Jensen:1983ac,Jensen:1988zx}. We assume that such a solution exists. Another notable instanton is the Hawking Moss one \cite{Hawking:1981fz}, where $\phi$ is a constant and equal to its value at the top of the potential barrier. One can think of it as a limit of the Coleman de Luccia instanton where the scalar field rolls a zero amount. There are also oscillating bounces \cite{Banks:2002nm, Hackworth:2004xb}, where the field configuration crosses the top of the potential barrier $n$ times, with $n>1$, instead of only once.

We can then discuss the one-loop determinant around this saddle. The quadratic action is again the one in \nref{qdacphi}, but with the background functions replaced by their sphere counterpart. We also assume that the fluctuation operator $M_{\phi}$ has a single negative mode, and that its zero modes are again only the ones associated with isometries that change the background scalar.

The difference is that the isometry generators $\xi^{a}$ are the ones of the sphere instead of flat space, with explicit form
\begin{equation}
\label{xitransl}
\xi_{a}=Y_{1}n_{a}+\sin \tau \, \cos \tau\,\n_{a}Y_{1}
\end{equation}
with $n^{a}$ the normal vector $\partial_{\tau}$. $Y_{1}$ is an $\ell=1$ harmonic in $S^{D-1}$, with the normalization such that $Y_{1}$ coincides with one of the embedding coordinates of $S^{D-1}$. For more details, see Appendix \nref{useid}. This normalization of $Y_{1}$ is the appropriate one such that the translation defined by $\xi^{a}$ reduces to the Cartesian translations, $\partial_{x}$, in the flat space limit.

Therefore, the integral over these generators is similar to the integral over spacetime positions for the center of the bounce in flat space, and gives a factor of $\text{Vol}(S^{D})$ with an appropriate Jacobian. Since locally the measure over translations in the sphere should be the same as in flat space, we can relate the field $\varphi=\xi^{a}\n_{a}\phi$ to a spacetime displacement as in \nref{flatsol0} and obtain again
\begin{equation}
\label{zphi0}
Z_{\phi,0}=\bigg(\frac{1}{2\pi D}\int \dot{\phi}^{2}\bigg)^{\frac{D}{2}}\text{Vol}(S^{D})=\bigg(\frac{(\xi^{a}\n_{a}\phi,\xi^{a}\n_{a}\phi)}{2\pi}\bigg)^{\frac{D}{2}}\text{Vol}(S^{D})
\end{equation}
with $\phi$ now evaluated in the sphere solution. We also rewrote the contribution in terms of the norm of $\xi^{a} \n_{a}\phi$ because this will be useful later in the dynamical gravity case. The norm is the same for all the $D$ linearly independent $\xi^{a}$ generators in \nref{xitransl}.

The sphere bounce contribution is therefore
\begin{equation}
\label{zphisphere}
Z_{\phi}(\text{bounce})=\int D\phi\,e^{-I}=(\pm i) \text{Vol}(S^{D})\bigg(\frac{1}{2 \pi D}\int \dot{\phi}^{2}\bigg)^{\frac{D}{2}}\text{Det}'|(-\n^{2}+v'')|^{-\frac{1}{2}}\,e^{-B-I_{\phi}(\phi_{\fv})}
\end{equation}

So, we see that this contribution has the same format as the flat space one, with the usual replacements of the background functions by their sphere counterparts. We can define a decay rate $\Gamma$ in terms of the ratio between \nref{zphisphere} and an appropriate false vacuum path integral in the sphere. This would lead to
\begin{equation}
\Gamma_{\text{QFT}}=\frac{1}{\text{Vol}(S^{D})}\text{Im}\bigg(\frac{Z_{\phi}(\text{bounce})}{Z_{\phi}(\phi_{\fv})}\bigg)=\bigg(\frac{1}{2 \pi D}\int \dot{\phi}^{2}\bigg)^{\frac{D}{2}}\frac{\text{Det}'|(-\n^{2}+v''(\phi))|^{-\frac{1}{2}}}{\text{Det}'(-\n^{2}+v''(\phi_{\fv}))^{-\frac{1}{2}}}\,e^{-B}
\end{equation}
where $Z_{\phi}(\phi_{\fv})$ is an pure false vacuum path integral in $S^{D}$. We interpret this $\Gamma$ as a decay rate in de Sitter space, which at least at the level of the action matches the decay rate discussed in \cite{Brown:2007sd}. If this classical solution $\phi$ interpolates smoothly to a flat space bounce as the radius of the sphere becomes big, we expect this $\Gamma_{\text{QFT}}$ to reduce to a decay rate in flat space.

\section{Vacuum decay with gravity}
\label{vacdecgr}

With gravity, the structure of the calculation is slightly different, namely, instead of having pure quantum field theory, we have a scalar that is minimally coupled to gravity, and the action of the system is
\begin{equation}
I_{E}=-\frac{1}{2\kappa^{2}}\int (R-2\Lambda)+\int\bigg(\frac{1}{2}g^{ab}\n_{a}\phi\n_{b}\phi+v(\phi)\bigg)
\end{equation}
with $\kappa^{2}=8 \pi G_{N}$, and we take the cosmological constant to be given by 
\begin{equation}
\label{cc}
\Lambda=\frac{(D-1)(D-2)}{2}
\end{equation}
where to avoid clutter, we picked the "radius of curvature" parameter in $\Lambda$ to be one. 

In part of this section, we will be interested in studying this system perturbatively in small $\kappa$, and we will assume that in an $\kappa$ expansion $v(\phi) \sim O(1)$. Note that we could have included the cosmological constant as part of the potential, even though it would be a part of the potential that becomes very big in the small $\kappa$ limit. To be specific, it would be a potential of the form
\begin{equation}
\label{potlambda}
V=\frac{(D-1)(D-2)}{2\kappa^{2}}+v(\phi)=\frac{\Lambda}{\kappa^{2}}+v(\phi)
\end{equation}

Therefore, in the $\kappa \rightarrow 0$ limit, one can think of this as a physical setup where the rolling of the scalar field $\phi$ does not change the potential $V$ too much compared to its offset value, which is very large and positive. As one takes $\kappa \rightarrow 0$, the first term in \nref{potlambda} makes the leading order classical metric be that of an Einstein manifold with radius of curvature one. It could, for example, be a sphere. Furthermore, the leading classical field configuration $\phi$ will be a solution to the scalar equations of motion in this Einstein manifold, as in section \nref{nobcksphere}. The backreaction from the scalar field can then be taken into account perturbatively in $\kappa^{2}$ to correct both the metric and the matter configuration of the classical solution.

The backgrounds we will be interested in will be Coleman de Luccia instantons \cite{Coleman:1980aw}, which are $SO(D)$ symmetric geometries where the scalar field $\phi$ rolls monotonically from one side of a potential barrier to the other. The line element is
\begin{equation}
ds^{2}=d\tau^{2}+a(\tau)^{2}d\Omega_{D-1}^{2}
\end{equation}
with $d\Omega_{D-1}^{2}$ the metric in the unit sphere $S^{D-1}$, and $a(\tau)$ vanish at both ends of the $\tau$ interval. The scalar field configuration is a profile $\phi(\tau)$, that satisfy the equation of motion
\begin{equation}
\ddot{\phi}+(D-1)H\dot{\phi}-V'(\phi)=0
\end{equation}
with $H=\frac{\dot{a}}{a}$, and where $a$ is fixed by the Friedman equation
\begin{equation}
\frac{(1-\dot{a}^{2})}{a^{2}}=\frac{2\kappa^{2}}{(D-1)(D-2)}\bigg(V-\frac{\dot{\phi}^{2}}{2}\bigg)=1+\frac{2\kappa^{2}}{(D-1)(D-2)}\bigg(v-\frac{\dot{\phi}^{2}}{2}\bigg)
\end{equation}

It should be easy to see that in the $\kappa \rightarrow 0$ limit, the solution to the Friedman equation is $a=\sin(\tau)$, where $\tau$ goes from $0$ to $\pi$, as it should be for the round sphere. The scalar solution then also reduces to the bounce in a sphere discussed in section \nref{nobcksphere}. 

Therefore, at small $\kappa$, the action of the Coleman de Luccia instanton reduces to the Einstein Hilbert action evaluated at a round sphere of radius one, plus the action of the scalar configuration at zero backreaction in the sphere geometry. This implies that, at the level of the action contribution, $e^{-I}$, the path integral over the Coleman de Luccia instanton is the same as the product of the pure gravity sphere path integral and the pure quantum field theory one in the sphere background, that is 
\begin{equation}
\label{actmatch}
e^{-I(\text{CdL})}=e^{-I_{\text{EH}}}e^{-I_{\phi}(\text{bounce})}(1+O(\kappa^{2}))
\end{equation}

We estimated the size of the error as follows: The difference between the exact solution for the instanton, let us call it $\Phi=(g_{ab},\phi)$, and the solution $\Phi_{0}=(g_{ab}(0),\phi_{0})$, that we find at zero backreaction, is $O(\kappa^{2})$. Therefore, since $\Phi$ is an exact saddle, deforming it to the zeroth order solution $\Phi_{0}$ will induce a change in the action of order $O(\kappa^{-2}\kappa^{4})=O(\kappa^{2})$. The $\kappa^{4}$ comes because the change in the action is quadratic in the deformation around the saddle, and the $\kappa^{-2}$ is the lowest power of $\kappa$ appearing in the action, more specifically in the Einstein-Hilbert term.

We will argue in section \nref{pertth} that the partition functions on the two sides of \nref{actmatch} also match at the one-loop level, up to corrections of order $\kappa^{2}$, in such a way that
\begin{equation}
Z(\text{CdL})=Z_{\text{GR}}(S^{D})Z_{\phi}({\text{bounce}})(1+O(\kappa^{2}))
\end{equation}
where we define $Z_{\text{GR}}(S^{D})$ to be the pure gravity path integral in the $S^{D}$ background, and $Z_{\phi}(\text{bounce})$ to be the pure scalar bounce path integral in $S^{D}$ that we discussed in subsection \nref{nobcksphere}.

In subsection \nref{flucop} we discuss how one would evaluate the one-loop determinant around the Coleman de Luccia instanton exactly. In subsection \nref{pertth}, we discuss how to compute the one-loop determinant perturbatively. The perturbative analysis reduces to studying how some zero modes of the solution at zero backreaction are lifted to light, non-zero modes, at small $\kappa$.

\subsection{One-loop fluctuation operator}
\label{flucop}

Taking an exact solution for the CDL instanton at non-zero $\kappa$ to be defined by a metric $g_{ab}$ and a background field configuration $\phi$, we now study fluctuations around it. This background with on-shell action $I_{E}$ will have a contribution to the path integral at the one-loop level of
\begin{equation}
\label{zcdldef}
Z(\text{CdL})=\frac{\int Dg D\phi\,e^{-I_{EH}-I_{\phi}}}{\int D\xi}=Z_{1-\text{loop}}e^{-I_{E}}
\end{equation}
with $Z_{\text{1-loop}}$ the one-loop determinant around the solution, and we used the path integral $\int D\xi$ in the denominator to indicate that we divide by the diffeomorphism group. The division over the gauge group has to be done at a more rigorous level through some BRST gauge fixing.

To find the one-loop determinant, we need to find the quadratic action of fluctuations around the solution $\delta_{2}I_{E}$ and do a path integral over them appropriately. For this, we need to define a measure for the path integral. We denote the fluctuations of the metric and scalar field, $\delta g_{ab}$ and $\delta \phi$ respectively, as
\begin{equation}
\delta g_{ab}=2\kappa h_{ab} ~~~~~~,~~~~~~\delta \phi=\varphi
\end{equation}
where the $2\kappa$ in the definition of $h_{ab}$ is such that the graviton kinetic term is canonically normalized. It is useful to think of both $h_{ab}$ and $\varphi$ as one collective fluctuation field $\Phi$ defined by
\begin{equation}
\label{bigphi}
\Phi=\begin{pmatrix}
    h_{ab}\\
    \varphi
\end{pmatrix}
\end{equation}
and the measure over the metric and fields can be thought of as a measure over $\Phi$. For this, we define the local norm
\begin{equation}
\label{locnorm}
(\Phi,\Phi)=\int (h_{ab}h^{ab}+\varphi^{2})
\end{equation}
and we take the measure over $\Phi$ to be the local measure derived from this norm, normalized such that
\begin{equation}
\label{Phimeas}
\int D\Phi e^{-\frac{1}{2}(\Phi,\Phi)}=\int Dh_{ab}e^{-\frac{1}{2}(h_{ab},h_{ab})}\int D\varphi\, e^{-\frac{1}{2}(\varphi,\varphi)}=1
\end{equation}

Note in the second equation of \nref{Phimeas} we made manifest that we can think of the local measure for $\Phi$ as a product over a measure for the metrics and one for the scalar. The measures factorize because $(h_{ab},0)$ and $(0,\varphi)$ are orthogonal in the inner product following from equation \nref{locnorm}. The scalar measure is normalized such that it matches the one we defined through equation \nref{sclocnorm}, and the gravity measure matches\footnote{We meant the measures have the same format, but they do become the same when $\kappa \rightarrow 0$. That is because the background geometry becomes $S^{D}$.} the one used in \cite{Anninos:2020hfj,Law:2020cpj}. 

Having chosen a local norm, one can write the quadratic action for the fluctuations, $\delta_{2}I_{E}$, as an inner product $\frac{1}{2}(\Phi,M_{0} \Phi)$ for an appropriate fluctuation operator $M_{0}$. If $M_{0}$ had no zero modes, doing the integral over $\Phi$ weighted by this action would be equivalent to evaluating a product over eigenvalues of $M_{0}$. However, the action of the theory is invariant under the diffeomorphisms
\begin{equation}
\label{gaugetr}
h_{ab}\rightarrow h_{ab}+\frac{1}{\sqrt{2}}(\n_{a}\xi_{b}+\n_{b}\xi_{a}) \text{ , } \varphi \rightarrow \varphi+\sqrt{2}\kappa \xi^{a}\n_{a}\phi
\end{equation}
where the factor of $\kappa$ in the $\varphi$ transformation appears because of the field redefinition in $h_{ab}$, and the $\sqrt{2}$ is convenient as discussed in \cite{Anninos:2020hfj,Law:2020cpj}. Heuristically, to make the path integral convergent, we have to cancel the integral over these zero modes in the numerator of equation \nref{zcdldef} with the division by the gauge group in the denominator, up to appropriate Jacobians. More rigorously, we need to do some gauge fixing, using, for example, a BRST construction.

To do the gauge fixing consistently, it is important to define a measure for the diffeomorphism path integral in equation \nref{zcdldef}, and we do so by picking a measure that follows from the local norm
\begin{equation}
(\xi^{a},\xi^{a})=\int \xi^{a}\xi_{a}
\end{equation}
and is normalized such that
\begin{equation}
\int D\xi\, e^{-\frac{1}{2}(\xi^{a},\xi^{a})}=1
\end{equation}

We implement the gauge fixing at a concrete level by adding a gauge fixing term $I_{gf}$ to the action, and an appropriate ghost determinant, consistent with the measures we picked for $\Phi$ and $\xi^{a}$. We insert the ghost determinant via a path integral over ghost fields $c$,$\bar{c}^{a}$ weighted by an appropriate ghost action $I_{gh}=\int(\bar{c},M_{gh}c)$. To be concrete, we pick a local norm for the ghosts of
\begin{equation}
(\bar{c}^{a},c_{a})=\int \bar{c}^{a}c_{a}
\end{equation}

We pick a measure over them that is normalized such that if we mode expand $c^{a}=\sum_{n}c_{n}f_{n}^{a}$ with $f_{n}^{a}$ a unit normalized basis function, then 
\begin{equation}
\int Dc=\int \prod dc_{n}\text{, with }~~~~\int dc_{n}\,c_{n}=1
\end{equation}

The gauge fixing, however, might not fix some special zero modes, which we treat separately. For example, diffeomorphisms associated with isometries. While they are important and must appear in the division by the gauge group, as stressed, for instance, in \cite{Donnelly:2013tia}, they do not change the background and cannot be gauge fixed via local gauge conditions in the metric and scalar. 

They form a residual gauge group $G$, and we denote their contribution to the division by diffeomorphisms in equation \nref{zcdldef} by $Z_{0}$. More explicitly
\begin{equation}
\label{ZGpi}
Z_{0}=\bigg(\int D\xi\bigg)_{\text{isometries}}^{-1}=\frac{1}{\text{Vol}(G)_{PI}}
\end{equation}
where we wrote $\text{Vol}(G)_{\text{PI}}$ to indicate the volume of this gauge group under the path integral measure. Note that $Z_{0}$ depends on the measure we choose for the $\xi$ path integral. However, as long as we define the ghost determinant consistently, the combination of $Z_{0}$ times the ghost determinant will not depend on the normalization of the measure.

Putting all these points together, the one-loop determinant can be found to be
\begin{equation}
Z_{1-\text{loop}}=Z_{0}\int D\Phi\,D'c\,D'\bar{c}\,e^{-\delta_{2}I_{E}-I_{gf}-I_{gh}}=Z_{0}\int D\Phi\,D'c\,D'\bar{c}\,e^{-\frac{1}{2}(\Phi,M\Phi)-(\bar{c}^{a},M_{gh}c^{a})}
\end{equation}
where we denoted the measure for the ghost as $D'c$ to mean we omit from their path integral the ghost zero modes, e.g, the residual gauge group generators.

Assuming $M$ to have a spectrum of eigenvalues $\lambda_{n}$ and the ghost operator $M_{gh}$ to have spectrum $\lambda_{gh,n}$, the one-loop determinant can be written as
\begin{equation}
\label{1loopeq}
Z_{1-\text{loop}}=Z_{0} \prod_{\lambda_{n}}\lambda_{n}^{-\frac{1}{2}} \prod_{\lambda_{gh,n}}'|\lambda_{gh,n}|
\end{equation}
where we took the absolute value of the ghost determinant, which intuitively follows because it is a Jacobian factor and therefore must be positive. It can be argued more rigorously that this is the correct prescription, see \cite{Polchinski:1988ua}. We again used the prime notation in $\prod'$ to mean we omit the zero modes of the ghost determinant from the product over eigenvalues. We also assumed that $M$ has no zero modes, but if it had, we would have included them in $Z_{0}$ as well.

We will now discuss how to compute all these factors for the Coleman de Luccia instanton. First, we find the quadratic action of fluctuations around the solution to be
\begin{equation}
\begin{gathered}
\label{delta2IE}
\delta_{2} I_{E}=\frac{1}{2}\int  \tilde{h}^{ab}(-\n^{2}h_{ab}+2\n_{(a}\n^{c}\tilde{h}_{b)c}-2\tensor{R}{_a^c_b^d}h_{cd}+2\kappa^{2}h_{c(a}\n_{b)}\phi\,\n^{c}\phi)\\
+\int\bigg[2\kappa \varphi(h^{ab}\n_{a}\n_{b}\phi+\n_{b}\tilde{h}^{ab}\n_{a}\phi)+\frac{1}{2}\varphi(-\n^{2}+v'')\varphi\bigg]
\end{gathered}
\end{equation}
where we defined
\begin{equation}
\tilde{h}_{ab}=h_{ab}-\frac{1}{2}g_{ab}h \text{, with} ~~~~ h=g^{ab}h_{ab}
\end{equation}

Because of the $\kappa$ in the definition of $h_{ab}$ the kinetic term for $h_{ab}$ is $O(1)$ in a $\kappa$ expansion, and the cross terms between the metric and scalar fluctuations are $O(\kappa)$. To gauge fix we add the following term to the action
\begin{equation}
I_{gf}=\frac{1}{2\sigma^{2}}\int P_{a}P^{a}=\int \bigg(\n^{b}\tilde{h}_{ab}-\gamma\kappa \varphi\n_{a}\phi\bigg)\bigg(\n_{c}\tilde{h}^{ac}-\gamma\kappa \varphi\n^{a}\phi\bigg)
\end{equation}
which corresponds to using the gauge constraint
\begin{equation}
\label{gaugecond}
P_{a}=\n^{b}\tilde{h}_{ab}-\gamma \kappa \varphi \n_{a}\phi
\end{equation}
with gauge fixing coefficient $\sigma=\frac{1}{\sqrt{2}}$. Identifying the gauge fixing coefficient is important to get a convenient normalization of the ghost determinant, as explained in Appendix \nref{gfnorm}. We introduced the convenient $\gamma$ term in equation \nref{gaugecond} because it will make the perturbation theory in section \nref{pertth} a bit simpler.

In adding $I_{gf}$ to the action, one obtains an overall action of
\begin{equation}
\begin{gathered}
\label{gfaction}
\delta_{2}I_{E}+I_{gf}=\frac{1}{2}\int  \tilde{h}^{ab}(-\n^{2}h_{ab}-2\tensor{R}{_a^c_b^d}h_{cd}+2\kappa^{2}h_{c(a}\n_{b)}\phi\,\n^{c}\phi)
\\+\int\bigg[2\kappa \varphi(h^{ab}\n_{a}\n_{b}\phi+(1-\gamma)\n_{b}\tilde{h}^{ab}\n_{a}\phi)+\frac{1}{2}\varphi(-\n^{2}+v''+2\gamma^{2}\kappa^{2}\dot{\phi}^{2})\varphi\bigg]
\end{gathered}
\end{equation}

Having worked this out, one can solve for the spectrum of fluctuations. For this, it is important first to have established a norm on the space of fluctuations as we did in equation \nref{locnorm}, so that we can write the action in the form
\begin{equation}
\delta_{2}I_{E}+I_{gf}=\frac{1}{2}(\Phi,M\Phi)
\end{equation}
with $M$ an appropriate self-adjoint fluctuation operator. Having this in hand, one can proceed to find the eigenvalues and eigenmodes of the fluctuation operator $M$, which consists of finding vectors $\Phi$ such that
\begin{equation}
M\Phi=\lambda \Phi
\end{equation}

From the local norm we introduced \nref{locnorm}, one can find the eigenvalue equation to be equivalent to
\begin{equation}
\begin{gathered}
\lambda \varphi=(-\n^{2}+v''+2\gamma^{2}\kappa^{2}\dot{\phi}^{2})\varphi+2\kappa(h^{ab}\n_{a}\n_{b}\phi+(1-\gamma)\n_{b}\tilde{h}^{ab}\n_{a}\phi)\\
\lambda \bigg(h_{ab}-\frac{1}{(D-2)}g_{ab}h\bigg)=(-\n^{2}h_{ab}-2\tensor{R}{_a^c_b^d}h_{cd}+2\kappa^{2}h_{c(a}\n_{b)}\phi\,\n^{c}\phi)\\
+2\kappa\bigg[\gamma \varphi \n_{a}\n_{b}\phi+\frac{(\gamma-1)}{2}(\n_{a}\phi \n_{b}\varphi+\n_{b}\phi \n_{a}\varphi)-\frac{1}{(D-2)}g_{ab}\varphi \n^{2}\phi\bigg]
\end{gathered}
\end{equation}

For later convenience, it is also useful to rewrite this eigenvalue equation as
\begin{equation}
\begin{gathered}
\label{flucgf}
\lambda \varphi=(-\n^{2}+v'')\varphi+2\kappa(h^{ab}\n_{a}\n_{b}\phi+\n_{b}\tilde{h}^{ab}\n_{a}\phi)-2\gamma \kappa \n^{a}\phi\,P_{a} \\
\lambda \bigg(h_{ab}-\frac{1}{(D-2)}g_{ab}h\bigg)=(-\n^{2}h_{ab}+2\n_{(a}\n^{c}\tilde{h}_{b)c}-2\tensor{R}{_a^c_b^d}h_{cd}+2\kappa^{2}h_{c(a}\n_{b)}\phi\,\n^{c}\phi)\\
-2\kappa\bigg[\frac{1}{2}(\n_{a}\phi \n_{b}\varphi+\n_{b}\phi \n_{a}\varphi)+\frac{1}{(D-2)}g_{ab}\varphi \n^{2}\phi\bigg]-2\n_{(a}P_{b)}
\end{gathered}
\end{equation}
with $P_{a}$ the gauge condition \nref{gaugecond}. In this form, one can read the fluctuation operator as one term coming purely from $\delta_{2}I_{E}$, and a term coming only from the gauge fixing term $I_{gf}$. 

An interesting observation is that the metric and scalar modes in \nref{flucgf} should decouple at very high energies. This is because the coupling between the metric and the scalar in the fluctuation operator \nref{flucgf} has only a single derivative, while the coupling of the metric and scalar with themselves has two, coming from their kinetic term. This implies that for very short wavelengths the solutions to the eigenvalue equation \nref{flucgf} decompose into almost pure scalar modes and almost pure metric modes.

However, at intermediate energy scales, the metric and scalar field are coupled, and we need to solve the coupled eigenvalue equation explicitly. In principle, one can solve this coupled system of equations numerically, but it can be quite cumbersome. We study the coupled system more explicitly for an instanton toy model in section \nref{toyinstsec}. However, for a generic scalar potential, we restrict ourselves to solutions with very small backreaction. 

As a part of the gauge fixing procedure, we also have to add a ghost term to the action, more explicitly
\begin{equation}
I_{gh}=\sigma^{-1}\int \bar{c}^{a}(-\delta_{\xi}P c)^{a}=\int \bar{c}^{a}(-\n^{2}c_{a}-\tensor{R}{_a^b}c_{b}+2\gamma \kappa^{2}\n_{a}\phi\,c^{b}\n_{b}\phi)=(\bar{c}^{a},M_{gh}c^{a})
\end{equation}

We introduced the factor of $\sigma^{-1}$ in the ghost action so that we do not need to worry about extra normalization factors later, as discussed in Appendix \nref{gfnorm}. This factor conveniently cancelled against a factor of $\frac{1}{\sqrt{2}}$ from our definition of the gauge transformation.

The eigenvalue equation for these modes is
\begin{equation}
\label{flucgh}
\lambda c^{a}=(-\n^{2}c^{a}-\tensor{R}{_a^b}c^{b})+2\gamma \kappa^{2} c^{b}\n_{b}\phi \n_{a}\phi
\end{equation}

The zero modes of \nref{flucgh} correspond to coordinate transformations that leave the gauge fixing term invariant. A good local gauge fixing condition will only have as zero modes coordinate transformations that leave the background itself invariant. For the Coleman de Luccia instanton, they are the generators of the rotation group of the spatial slices $S^{D-1}$, e.g, $SO(D)$.

These ghost zero modes that we do not integrate over can be thought of equivalently as being the contribution $Z_{0}$ we defined in \nref{ZGpi}. To evaluate $Z_{0}$, we note that the measure over diffeomorphisms associated to the residual gauge group $G$ should be $G$ invariant. Also, since it is also a group measure over $G$, it should be equal to the $G$-invariant group measure up to a normalization factor. 

To determine the normalization factor, one has to relate the path integral measure to the canonical group measure, using the ideas, for example, from \cite{Anninos:2020hfj,Law:2020cpj}. In doing so, we conclude that the contribution from these isometries is 
\begin{equation}
\label{z0}
Z_{0}=\frac{1}{\text{Vol}(SO(D))_{c}}\bigg(\frac{2\pi(\sqrt{2}\kappa)^{2}}{(\xi^{a},\xi^{a})}\bigg)^{\frac{|SO(D)|}{2}}
\end{equation}
with $\text{Vol}(SO(D))_{c}$ the canonical group volume of $SO(D)$, and $\xi^{a}$ one of the $SO(D)$ generators that satisfy the usual commutation relations. The norm $(\xi^{a},\xi^{a})$ is the same for all these generators.

Note that for the instanton at $\kappa=0$ the operator $M_{gh}$ will only contain the gravitational piece, coming from varying $\n^{b}\tilde{h}_{ab}$. Therefore, it will have as zero modes the entire sphere isometry group, $SO(D+1)$.

\subsection{Perturbation theory}
\label{pertth}

Suppose the backreaction from the scalar is very small. In that case, the metric will be very close to the metric in $S^{D}$, which we call $g_{ab}(0)$, and the field configuration will reduce to its zero backreaction limit as in section \nref{nobcksphere}, which we call $\phi(0)$. That is,
\begin{equation}
g_{ab}=g_{ab}(0)+\kappa^{2}H_{ab} ~~~~~,~~~~~  \phi=\phi(0)+\kappa^{2}\Delta \phi
\end{equation}

Both $H_{ab}$ and $\Delta \phi$ are $O(1)$, but their exact form is a power series in $\kappa^{2}$. 

We are then interested in studying the spectrum of $M$ and $M_{\text{gh}}$, e.g, in solving \nref{flucgf} and \nref{flucgh}, perturbatively. We will start with the analysis of $M$, and the analysis of the ghost operator $M_{\text{gh}}$ will follow similarly. 

As one takes $\kappa \rightarrow 0$, the fluctuation operator $M$ of the gravity+scalar system will break down into approximately the direct sum of a pure gravity fluctuation operator $M_{GR}$ and a pure scalar fluctuation operator $M_{\phi}$, with both operators evaluated at the leading background configuration $(g_{ab}(0),\phi(0))$. More explicitly
\begin{equation}
M=\begin{pmatrix}
     M_{GR} & 0\\
     0 & M_{\phi}
\end{pmatrix}+\begin{pmatrix}O(\kappa^{2}) & O(\kappa)\\
O(\kappa) & O(\kappa^{2})\end{pmatrix}
\end{equation}

The spectrum of fluctuations around the sphere in the de Donder gauge, e.g, $M_{GR}$, is well-known \cite{Polchinski:1988ua}, and the general structure of the scalar fluctuation operator in the sphere background, $M_{\phi}$, can be computed from quantum field theory alone, as in section \nref{nobcksphere}. So, perturbatively it must be that the eigenvalues of $M$ are divided into almost pure gravity modes $\lambda_{GR}$, that are almost eigenvalues of $M_{GR}$, and almost pure scalar modes $\lambda_{\phi}$, that are almost eigenvalues of $M_{\phi}$. The eigenvalues and eigenmodes thus divide as
\begin{equation}
\begin{gathered}
\label{appeignm}
\lambda_{GR}=\lambda_{GR}^{(0)}+O(\kappa^{2}) \text{    ,    } \Phi_{GR}=\begin{pmatrix}h_{ab}^{(0)}+O(\kappa^{2})\\O(\kappa) \end{pmatrix} \\
\lambda_{\phi}=\lambda_{\phi}^{(0)}+O(\kappa^{2}) \text{    ,    } \Phi_{\phi}=\begin{pmatrix}O(\kappa)\\\varphi^{(0)}+O(\kappa^{2}) \end{pmatrix} 
\end{gathered}
\end{equation}

The $\lambda_{GR}^{(0)}$ and $\lambda_{\phi}^{(0)}$ are eigenvalues of $M_{GR}$ and $M_{\phi}$ respectively, with $h_{ab}^{(0)}$ and $\varphi^{(0)}$ their respective eigenmodes\footnote{Strictly speaking, \nref{appeignm} applies in the absence of degeneracies between coupled gravitational and scalar modes. Degenerate subspaces require degenerate perturbation theory and can exhibit $O(\kappa)$ eigenvalue splittings, while their combined determinant still has no linear correction.}. This implies that for any non-zero modes, the backreaction correction to the eigenvalues is only perturbatively small and can be neglected at leading order. This is also true for the non-zero eigenvalues of the ghost fluctuation operator $M_{gh}$.

A possible loophole in this approximation is that while the corrections in \nref{appeignm} are suppressed by $\kappa^{2}$, they could in principle increase with energy and become relevant for small enough mode wavelengths. However, at very high energies, the fluctuation operator is dominated by the Laplacians, so any correction to the eigenvalues comes from a correction to the Laplacian. These corrections can scale with energy, but they will be multiplicative corrections in $\kappa^{2}$, so the eigenvalues change as $\lambda=\lambda^{(0)}(1+O(\kappa^{2}))$ and the effect is still subleading. The net effect of these corrections should also be small, if we perform the determinant comparison using a common regulator and consistent subtraction of local divergences.

However, the backreaction can lift zero modes, and therefore changes their leading order value to something small but finite. We call these lifted zero modes "light modes". This is a relevant point, because it means that they contribute to the one-loop determinant via the product over non-zero eigenvalues in \nref{1loopeq}, instead of the integral over zero modes. 

Having considered all of these points, we can write $Z(\text{CdL})$ at leading order in perturbation theory up to some unknown factors we discuss later. The $e^{-I}$ contribution to $Z(\text{CDL})$ matches that of the product $Z_{\text{GR}}(S^{D})Z_{\phi}(\text{bounce})$, as we argued in equation \nref{actmatch}. Then, the product of non-zero eigenvalues in $Z(\text{CdL})$, e.g, in equation \nref{1loopeq}, is composed of light modes, and the non-light eigenvalues of $M_{GR}$, $M_{\phi}$, and $M_{gh}$. These non-light eigenvalues all appear as contributions to the one-loop determinant of either $Z_{\phi}(\text{bounce})$ or $Z_{\text{GR}}(S^{D})$, and therefore
\begin{equation}
Z'(\text{CdL})=Z_{\text{light}}Z'_{\text{GR}}(S^{D})Z'_{\phi}(\text{bounce})(1+O(\kappa^{2}))
\end{equation}
with $Z_{\text{light}}$ the appropriate product over light eigenvalues, and $Z'$ is the notation we introduced in equation \nref{zprime}. Therefore, the only part that differs between these path integrals is that $Z(\text{CdL})$ has light mode contributions and its own zero mode contribution $Z_{0}$, while $Z_{\phi}(\text{bounce})$ and $Z_{\text{GR}}(S^{D})$ have zero mode contributions $Z_{\phi,0}$ and $Z_{\text{GR},0}$. Thus, we can conclude that
\begin{equation}
Z(\text{CdL})=Z_{\phi}(\text{bounce})Z_{\text{GR}}(S^{D})\bigg(\frac{Z_{0}\prod_{\text{light}}\lambda_{n}^{-\frac{1}{2}}\prod_{\text{light}}|\lambda_{gh}|}{Z_{\phi,0}Z_{\text{GR,0}}}\bigg)(1+O(\kappa^{2}))
\end{equation}
where in the numerator we used $\prod_{\text{light}}$ to mean the product over light eigenvalues, and we included both the contributions from the matter fluctuation operator $M$ and the ghost operator $M_{gh}$. To discuss what these light modes are, we first discuss the zero mode contributions to the path integrals. 

As we discussed in stating equation \nref{z0}, the Coleman de Luccia instantons have exact zero modes that are associated with the preserved isometry group $SO(D)$, which leave both the metric and the field configuration $\phi$ invariant. The zero modes contribution of the pure gravity sphere path integral $Z_{GR}(S^{D})$ has the same form as $Z_{0}$, but the difference is that the norm $(\xi^{a},\xi^{a})$ is computed in the zero backreaction geometry $S^{D}$, and its isometry group is $SO(D+1)$. One can think of $SO(D+1)$ as being composed of the $SO(D)$ generators and $D$ extra generators that can be viewed as translations along $S^{D}$. Therefore,  
\begin{equation}
\begin{gathered}
Z_{GR,0}(S^{D})=\frac{1}{\text{Vol}(SO(D+1))_{c}}\bigg(\frac{2\pi (\sqrt{2}\kappa)^{2}}{(\xi^{a},\xi^{a})_{\kappa=0}}\bigg)^{\frac{|SO(D+1)|}{2}}\\
=\frac{1}{\text{Vol}(SO(D))_{c}\times \text{Vol}(S^{D})}\bigg(\frac{2\pi (\sqrt{2}\kappa)^{2}}{(\xi^{a},\xi^{a})_{\kappa=0}}\bigg)^{\frac{|SO(D)|+D}{2}}
\end{gathered}
\end{equation}

This is similar in form to the contribution \nref{z0}, but it has extra factors of $\kappa$ because, as we discussed, $|SO(D+1)|=|SO(D)|+D$. The canonical volume factor is also similar but has an extra factor of $\frac{1}{\text{Vol}(S^{D})}$ because $\text{Vol}(SO(D+1))_{c}=\text{Vol}(SO(D))_{c}\text{Vol}(S^{D})$. 

One can think of the difference in symmetry factors as the fact that the Coleman de Luccia solution is a deformed sphere, so it has a preferred center. Meanwhile, in the round sphere, all centers are the same, so we have an extra symmetry factor of $\frac{1}{\text{Vol}(S^{D})}$. Interestingly, the contribution of the pure scalar path integral $Z_{\phi,0}$ we found in \nref{zphi0} has a factor of $\text{Vol}(S^{D})$ which cancels this difference, and the canonical group volume factors of $Z_{0}$ match those of $Z_{\text{GR},0}Z_{\phi,0}$. 

The reasoning is that even with no backreaction, the symmetry group of the sphere+scalar field configuration is $SO(D)$, since shifting the sphere center does not change the sphere, but it does affect the scalar field. Alternatively, while in $Z_{GR,0}$ we are dividing by translations, in $Z_{\phi,0}$ we are integrating over them, so these contributions should cancel up to Jacobian factors. 

Using the zero mode contribution of the pure scalar path integral $Z_{\phi,0}$ we found in \nref{zphi0} and putting it all together explicitly, we can compute the ratio of the zero mode contributions at leading order in $\kappa$ to be
\begin{equation}
\begin{gathered}
\label{z0ratio}
\frac{Z_{0}}{Z_{GR,0}Z_{\phi,0}}=\bigg(\frac{(\xi^{a},\xi^{a})_{\kappa=0}}{(\sqrt{2}\kappa)^{2}(\xi^{a}\n_{a}\phi,\xi^{a}\n_{a}\phi)_{\kappa=0}}\bigg)^{\frac{D}{2}}(1+O(\kappa^{2}))
\end{gathered}
\end{equation}
where we replaced \nref{z0} by its leading order value in $\kappa$, so that we could cancel it with the $SO(D)$ zero mode integral in $Z_{GR,0}$.

We will now discuss the contribution of the light modes $Z_{\text{light}}$ and explain why they cancel \nref{z0ratio} at leading order. First, we discuss what the light modes are. They divide into light modes of $M$ and light modes of $M_{\text{gh}}$, so let us start by discussing the light modes of $M$.

The pure gravity fluctuation operator in the de Donder gauge, $M_{GR}$, has no zero modes for the sphere \cite{Polchinski:1988ua}, so their correction is not relevant. For a generic scalar background in a fixed geometry however, $M_{\phi}$ will have zero modes associated with isometries that change $\phi$ non-trivially as we discussed in section \nref{nobcksphere}. We furthermore assume these are the only zero modes of $M_{\phi}$. Therefore, the only light modes of $M$ come from these zero modes.

Note that these light modes correspond precisely to the isometries broken by the backreaction of the scalar, and therefore correspond to the zero modes which appeared in $Z_{GR,0}Z_{\phi,0}$ but not in $Z_{0}$. In other words, these modes are still running in the path integral, but instead of being treated as exact zero modes, they are contributing as non-zero modes taken into account in $Z_{\text{light}}$.

There is a similar statement for the ghost zero modes. Namely, in the zeroth order background, the zero modes of the ghost operator are the $SO(D+1)$ generators, while in the instanton, the exact zero modes are the generators of $SO(D)$. So, it must be that the same $|SO(D+1)|-|SO(D)|=D$ broken isometries give rise to ghost light modes.

The only thing left is to check what the eigenvalues of these light modes are, concretely. We start by studying the explicit form of the light modes of $M$. Our strategy for finding these light modes will consist of motivating an Ansatz for their eigenmode in equation \nref{PhiAnsatz}, and showing that it solves the eigenvalue equation at the relevant order of perturbation theory.

As a reminder, these light modes come from zero modes of $M_{\phi}$ of the form
\begin{equation}
\label{zerothphi}
\varphi=\mathcal{L}_{\xi}\phi_{0} ~~~~,~~~~ h_{ab}=0
\end{equation}
with $\phi_{0}$ the classical field configuration $\phi$ at zero backreaction, for example discussed in section \nref{nobcksphere}. $\mathcal{L}_{\xi}$ is the Lie derivative along an isometry $\xi^{a}$ of the form \nref{xitransl}. Using that \nref{zerothphi} is an eigenmode of \nref{flucgf} at zeroth order in $\kappa$, we can find its form to all orders in $\kappa$ by imposing that \nref{flucgf} holds. 

We note that \nref{zerothphi} is a coordinate transformation at zeroth order, since $\xi^{a}$ is an isometry of the round sphere. Therefore, if there were no gauge fixing term in \nref{flucgf}, a solution to the eigenvalue equation that reduces to \nref{zerothphi} at $\kappa \rightarrow 0$ would have been the exact coordinate transformation of the background induced by $\xi^{a}$, that is
\begin{equation}
\label{fullphi}
h_{ab}=\frac{1}{2\kappa}\mathcal{L}_{\xi}(g_{ab}(0)+\kappa^{2}H_{ab})=\frac{1}{2}\kappa \mathcal{L}_{\xi}H_{ab} ~~~~~,~~~~~ \varphi=\mathcal{L}_{\xi}\phi
\end{equation}

The reason is that \nref{fullphi} manifestly reduces to \nref{zerothphi} at $\kappa=0$, and it would be an exact eigenmode of the version of \nref{flucgf} without the gauge fixing term, since it is a coordinate transformation. Note that we used that $\xi^{a}$ is an isometry of the zeroth order background $g_{ab}(0)$. Also, the $(2\kappa)^{-1}$ is present in the $h_{ab}$ transformation in \nref{fullphi} because we picked the coordinate transformation such that $\varphi$ changes in the specified way.  

It is then natural to think of the exact eigenmode following from the zeroth order mode \nref{zerothphi} as being \nref{fullphi} plus an extra term induced by the gauge fixing term in \nref{flucgf}, that is
\begin{equation}
\label{Ansatzpert}
\lambda=\kappa^{2}\lambda_{1}+O(\kappa^{4}) ~~~~,~~~~ \Phi=\begin{pmatrix}
   \frac{1}{2} \kappa\mathcal{L}_{\xi}H_{ab}+ \kappa h_{ab}^{(1)}+O(\kappa^{3})\\ \mathcal{L}_{\xi}\phi+O(\kappa^{2})
\end{pmatrix}
\end{equation}

We can then fix $h_{ab}^{(1)}$ using \nref{flucgf}. More specifically, we fix $h_{ab}^{(1)}$ by using that the left-hand side of the metric eigenvalue equation in \nref{flucgf} is $O(\kappa^{3})$, and therefore the $O(\kappa)$ contribution from the right-hand side must be zero. The strategy to find $h_{ab}^{(1)}$ is then the following: The \nref{fullphi} mode is an exact coordinate transformation, so it gives a zero contribution to the right-hand side of \nref{flucgf} except for its contribution to the gauge fixing term $\n_{(a}P_{b)}$. Therefore, the $h_{ab}^{(1)}$ contribution has to cancel this contribution at $O(\kappa)$. The simplest way this can happen is if we pick $h_{ab}^{(1)}$ to come from a coordinate transformation as well, such that its only contribution to the right-hand side of \nref{flucgf} would also be through the gauge fixing term. In fact, we can make $(\kappa h_{ab}^{(1)},0)$ into an exact coordinate transformation of the instanton by adding an $O(\kappa^{2})$ term to the $\varphi$ component. This is allowed since this addition can be absorbed in the $O(\kappa^{2})$ error of \nref{Ansatzpert}, so that if $h_{ab}^{(1)}=2\n_{(a}v_{b)}$ then
\begin{equation}
\begin{gathered}
\label{PhiAnsatz}
\Phi=\begin{pmatrix}\frac{1}{2}\kappa \mathcal{L}_{\xi}H_{ab}\\ \xi^{a}\n_{a}\phi\end{pmatrix}+\begin{pmatrix}2\kappa\n_{(a}v_{b)}\\0\end{pmatrix}+O(\kappa^{2})
=\begin{pmatrix}\frac{1}{2}\kappa \mathcal{L}_{\xi}H_{ab}\\\xi^{a}\n_{a}\phi\end{pmatrix}+\kappa\begin{pmatrix} \mathcal{L}_{v}g_{ab}\\2\kappa\mathcal{L}_{v}\phi\end{pmatrix}+O(\kappa^{2})
\end{gathered}
\end{equation}

Then, if we can solve \nref{flucgf} at $O(\kappa)$ with this Ansatz, this is the correct solution for $h_{ab}^{(1)}$ at this order. For the solution to exist, we just need that $\n_{(a}P_{b)}$ is zero at $O(\kappa)$. $P_{a}$ itself is already $O(\kappa)$, more explicitly
\begin{equation}
P_{a}=\kappa p_{a}=\kappa(\n^{b}\big(\frac{1}{2}\mathcal{L}_{\xi}\tilde{H}_{ab}+\tilde{h}_{ab}^{(1)}\big)-\gamma \varphi \n_{a}\phi+O(\kappa^{2}))
\end{equation}
which implies that $\n_{(a}p_{b)}$ has to vanish at $O(1)$. Since this is an equality of $O(1)$ terms, we can replace all background functions, such as covariant derivatives and functions of $\phi$, by their zeroth order saddle value. 

The operator $(O u)_{ab}=2\n_{(a}u_{b)}$ from vectors to symmetric two-tensors has kernel by definition equal to the isometry group of the manifold. Since we can replace the covariant derivatives by the sphere ones at this order, the isometry group is $SO(D+1)$. Therefore, $(Ou)=0$ implies that $u^{a}$ is along an isometry direction of $S^{D}$. The only isometry consistent with the modes we are discussing is the isometry $\xi^{a}$ itself, therefore
\begin{equation}
p^{a}=A \xi^{a}+O(\kappa^{2})
\end{equation}
with $A$ a constant fixed by requiring the two sides to match at $O(1)$. To be more explicit,
\begin{equation}
\label{Aeq}
A=\bigg(\frac{\int  \xi^{a}p_{a}}{\int \xi^{a}\xi_{a}}\bigg)_{\kappa=0}=-\gamma \bigg(\frac{\int (\xi^{a}\n_{a}\phi)^{2}}{\int \xi^{a}\xi_{a}}\bigg)_{\kappa=0}=-\gamma \bigg(\frac{(\xi^{a}\n_{a}\phi,\xi^{a}\n_{a}\phi)}{(\xi^{a},\xi^{a})}\bigg)_{\kappa=0}
\end{equation}
where we used the fact that any vector $p^{a}$ on the sphere can be expanded into a set of eigenmodes of the Laplacian $-\bar{\n}^{2}=-(\n^{2})_{\kappa=0}$ on the sphere, and $\xi^{a}$ is one of them. Therefore, one can find $A$ by projecting $p^{a}$ along the $\xi^{a}$ direction. Then, we can solve for $h_{ab}^{(1)}$ by solving for the coordinate transformation $v^{a}$ such that $p^{a}-A\xi^{a}=O(\kappa^{2})$.

More explicitly, if $\bar{\n}=(\n)_{\kappa=0}$ is the covariant derivative in $S^{D}$ and $p_{0}^{a}=(p^{a})_{\kappa=0,h_{ab}^{(1)}=0}$, we need to solve
\begin{equation}
\label{gagach}
(\bar{\n}^{2}+(D-1))v^{a}=A\xi^{a}-p_{0}^{a}
\end{equation}

To analyze whether this equation has solutions, note that the kernel of $(\bar{\n}^{2}+(D-1))$ as a vector operator is the space of isometries of $S^{D}$. However, the right-hand side of \nref{gagach} is orthogonal to this space by construction\footnote{Note that the $h_{ab}^{(1)}$ term does not contribute to \nref{Aeq}.}; therefore, one can always solve for an $O(1)$ $v^{a}$ such that the equation holds, and the Ansatz for the mode was correct. 

It is then simple to evaluate the eigenvalue shift at leading order $\lambda_{1}$ by noticing that it is the same as evaluating the action at this order, namely
\begin{equation}
\frac{1}{2}(\Phi,\lambda \Phi)=\frac{1}{2}\kappa^{2}\lambda_{1} (\varphi,\varphi)_{\kappa=0}+O(\kappa^{4})=\delta_{2}I_{E}+I_{gf}
\end{equation}

We can evaluate the action as the sum of the contribution from the second-order variation, $\delta_{2}I_{E}$, and the contribution from the gauge fixing term $I_{gf}$. The contribution from $\delta_{2}I_{E}$ has zero action at order $\kappa^{2}$, because the mode is an exact coordinate transformation up to corrections of $O(\kappa^{2})$. Therefore, this error can only shift the action to $O(\kappa^{4})$, which can be ignored at the order we care about. This implies that the action at $O(\kappa^{2})$ comes from $I_{gf}$ alone. Since $I_{gf}\sim \kappa^{2}\int p_{a}p^{a}$, to evaluate it at $O(\kappa^{2})$ we only need $p_{a}$ at zeroth order in $\kappa$. Therefore

\begin{equation}
\begin{gathered}
I=\frac{1}{2}\kappa^{2}\lambda_{1} (\varphi,\varphi)+O(\kappa^{4})=\kappa^{2}\bigg(\int p^{a}p_{a}\bigg)_{\kappa=0}+O(\kappa^{4})\Rightarrow \lambda_{1}=2\gamma^{2} \frac{(\xi^{a}\n_{a}\phi,\xi^{a}\n_{a}\phi)_{\kappa=0}}{(\xi^{a},\xi^{a})_{\kappa=0}}
\end{gathered}
\end{equation}
which allows us to conclude more explicitly that
\begin{equation}
\label{shiftm}
\lambda=\kappa^{2}\lambda_{1}+O(\kappa^{4})=2\gamma^{2}\kappa^{2} \bigg(\frac{(\xi^{a}\n_{a}\phi,\xi^{a}\n_{a}\phi)}{(\xi^{a},\xi^{a})}\bigg)_{\kappa=0}+O(\kappa^{4})
\end{equation}

So, as promised, the shift in the eigenvalue is positive definite and has a simple, explicit form in terms of the background $\phi$ at leading order in the backreaction. We see that the eigenvalue would be zero at this order if $\gamma$ were zero, so that is the reason why the $\gamma$ term is convenient. 

The light modes left to discuss explicitly are those of the ghost operator $M_{gh}$. We already discussed that these light modes correspond to the broken isometries $\xi^{a}$, which are in $SO(D+1)$ but not $SO(D)$, so we can evaluate the eigenvalue shift using simple first-order perturbation theory. That is, since we know the eigenmodes are $\xi^{a}$ at zeroth order, if the ghost operator $M_{gh}$ is deformed by $\kappa^{2}\delta M_{gh}$ from its zeroth order value, then the eigenvalue shift is
\begin{equation}
\begin{gathered}
\label{ghostshift}
(u^{a},M_{gh} u^{a})=\kappa^{2}(\xi_{a},\delta M_{gh}\xi_{a})+O(\kappa^{4})=\int \xi^{a}(-\kappa^{2}\n^{b}\mathcal{L}_{\xi}\tilde{H}_{ab}+2 \gamma \kappa^{2}\xi^{b}\n_{b}\phi\, \n_{a}\phi)+O(\kappa^{4})\\
=2\gamma \kappa^{2}\int \varphi^{2}+O(\kappa^{4})=\kappa^{2}\lambda_{gh}^{(1)}(\xi_{a},\xi_{a})+O(\kappa^{4})
\end{gathered}
\end{equation}
where we assumed that the exact eigenmode following from $\xi^{a}$ is $u^{a}=\xi^{a}+O(\kappa^{2})$. Note that we simplified \nref{ghostshift} by integrating by parts and using that $\n_{a}\xi_{b}$ for the broken isometries is antisymmetric at $O(1)$, so we could get rid of the $\n^{b}\mathcal{L}_{\xi}\tilde{H}_{ab}$ term. Therefore, the eigenvalue of these previously ghost zero modes is
\begin{equation}
\label{shiftgh}
\lambda_{gh}=2\gamma \kappa^{2}\bigg(\frac{(\xi^{a}\n_{a}\phi,\xi^{a}\n_{a}\phi)}{(\xi^{a},\xi^{a})}\bigg)_{\kappa=0}+O(\kappa^{4})
\end{equation}

This, therefore, concludes the perturbation theory discussion for the light modes. We can then compute how they contribute to the one-loop determinant in \nref{1loopeq}. Namely, since we have $D$ of these modes, they will contribute as
\begin{equation}
\label{zlight}
Z_{\text{light}}=(|\lambda_{gh}|\lambda^{-\frac{1}{2}})^{D}=\bigg(2\kappa^{2}\bigg(\frac{(\xi^{a}\n_{a}\phi,\xi^{a}\n_{a}\phi)}{(\xi^{a},\xi^{a})}\bigg)_{\kappa=0}\bigg)^{\frac{D}{2}}(1+O(\kappa^{2}))
\end{equation}
which cancels \nref{z0ratio} at leading order. Also, note that the factors of $\gamma$ canceled, as they should, since the one-loop determinant should be gauge fixing invariant. We can thus finally conclude that
\begin{equation}
\label{ZCdLrev}
Z(\text{CdL})=Z_{\text{GR}}(S^{D})Z_{\phi}(\text{bounce})(1+O(\kappa^{2}))
\end{equation}
as we wanted to find. An interesting consequence of \nref{ZCdLrev} is that, for $D \geq 4$, the phase of $Z(\text{CdL})$ is the same as that of the pure-gravity sphere partition function in lower dimension, $Z_{\mathrm{GR}}(S^{D-1})$. This is suggestive, since in the thin-wall limit the bubble wall has a $dS_{D-1}$ worldvolume.

Another possible, but less rigorous, way we could have obtained the light modes contribution $Z_{\text{light}}$ is by using the fact that the light modes are a coordinate transformation at leading order to cancel them out directly with the volume of the diffeomorphism group they correspond to, up to some Jacobian factors.

That is, for each of these modes acting with a small diffeomorphism of size $\Delta \alpha$ along the relevant direction, as $\Delta \alpha \xi^{a}$, should change $\varphi$ by $\Delta \alpha \sqrt{2}\kappa \xi^{a}\n_{a}\phi$. Therefore, the ratio of the path integral over one of these modes and its associated diffeomorphism is approximately
\begin{equation}
\begin{gathered}
\bigg(\frac{\int D\Phi}{\int_{\text{}} D\xi }\bigg)_{\text{translation}} \approx \bigg(\frac{\Delta \alpha}{\sqrt{2\pi}}\sqrt{2}\kappa (\xi^{a}\n_{a}\phi,\xi^{a}\n_{a}\phi)_{\kappa=0}^{\frac{1}{2}}\bigg)\bigg(\frac{\Delta \alpha}{\sqrt{2\pi}} (\xi^{a},\xi^{a})_{\kappa=0}^{\frac{1}{2}}\bigg)^{-1}\\=\bigg(\frac{(\sqrt{2}\kappa)^{2}(\xi^{a}\n_{a}\phi,\xi^{a}\n_{a}\phi)_{\kappa=0}}{(\xi^{a},\xi^{a})_{\kappa=0}}\bigg)^{\frac{1}{2}}
\end{gathered}
\end{equation}
which precisely matches the single broken isometry contribution of \nref{zlight} at leading order. 

\subsection{Final result}

Having established the result for the Coleman de Luccia path integral at small backreaction in section \nref{pertth}, we can discuss its decay rate interpretation. To do so, one must define the decay rate for the system in the presence of gravity. We use the natural definition following from the quantum field theory one in \nref{decrate}, adapted to our prescription in section \nref{presc}. Namely, we define
\begin{equation}
\label{decrategr}
\Gamma=\frac{1}{\text{Vol}(S_{\fv}^{D})}\text{Im}\bigg(\frac{Z(\text{CdL})}{Z_{\text{fv}}(S^{D})}\bigg)
\end{equation}
where we define $Z_{\text{fv}}(S^{D})$ to be the gravity+matter path integral around the solution with $\phi=\phi_{\text{fv}}$ everywhere, and with a round sphere geometry. The difference of this geometry from the $S^{D}$ we discussed before is that the round sphere here will be sourced by an effective cosmological constant given by the sum of \nref{cc} and an extra term coming from the constant scalar potential $v(\phi_{\fv})$. $\text{Vol}(S_{\fv}^{D})$ is the volume of this round sphere. At the level of the action contribution, this definition of decay rate also follows from the principle of detailed balance \cite{Banks:2002nm}.

Since $Z_{\fv}(S^{D})$ is a path integral in the same theory as the Coleman de Luccia instanton, its quadratic action has the same format as \nref{delta2IE}, with the background functions in the formula appropriately modified. The saddle of pure false vacuum is a perfect round sphere with a constant background scalar, so the one-loop determinant of the metric and the scalar exactly factorizes. 

Since there are no light modes to worry about in this saddle, and backreaction from the constant scalar potential $v(\phi_{\fv})$ is subleading, we can simplify $Z_{\fv}(S^{D})$. Namely, we can approximate $Z_{\fv}(S^{D})$ by the pure gravity path integral in $S^{D}$, $Z_{\text{GR}}(S^{D})$, and the quantum field theory path integral $Z_{\phi}(\phi_{\fv})$ around the false vacuum configuration in $S^{D}$. That is
\begin{equation}
Z_{\fv}(S^{D})=Z_{\text{GR}}(S^{D})Z_{\phi}(\phi_{\fv})(1+O(\kappa^{2}))
\end{equation}

We can then conclude that
\begin{equation}
\frac{Z(\text{CdL})}{Z_{\text{fv}}(S^{D})}=\frac{Z_{\text{GR}}(S^{D})Z_{\phi}(\text{bounce})}{Z_{\text{GR}}(S^{D})Z_{\phi}(\phi_{\fv})}(1+O(\kappa^{2}))=\frac{Z_{\phi}(\text{bounce})}{Z_{\phi}(\phi_{\fv})}(1+O(\kappa^{2}))
\end{equation}

Therefore, since $\text{Vol}(S_{\fv}^{D})=\text{Vol}(S^{D})(1+O(\kappa^{2}))$, the definition of decay rate \nref{decrategr} matches the pure quantum field theory one in \nref{decrate} at small backreaction. More precisely, it matches a de Sitter version of that decay rate. 

Note that if we interpreted the formula \nref{decrategr} as coming from a Callan and Coleman argument, as in \cite{Callan:1977pt}, then the original decay rate definition that leads to \nref{decrategr} would be slightly different. Namely, it would have an extra factor of $2$ in front of it, which would need to cancel against an additional factor of $\frac{1}{2}$ in front of $Z(\text{CdL})$, to end up effectively as \nref{decrategr}. This factor of $\frac{1}{2}$ would have to come from some subtle negative mode prescription, such as the one discussed in \cite{Callan:1977pt}. In section \nref{flatsol}, we introduced this factor of $\frac{1}{2}$ by hand in front of the $Z_{\phi}(\text{bounce})$ term in the decay rate. However, this was just a trick so that we could write $Z_{\phi}(\text{bounce})$ throughout the paper without the extra factor of $\frac{1}{2}$ in its definition.

From a more fundamental point, this factor of $\frac{1}{2}$ is physical and should have been included as part of the definition of $Z_{\phi}(\text{bounce})$. Therefore, this seems to imply that at least one of the negative modes of the gravity+matter path integral $Z(\text{CdL})$, done correctly, would have the subtle factor of $\frac{1}{2}$ discussed in \cite{Callan:1977pt}. This seems to be a counterintuitive point about the gravitational path integral that might not have been obvious a priori.

One might wonder if other negative modes in gravity should also come accompanied by such factors of $\frac{1}{2}$. However, we should point out that these additional negative mode corrections would not affect the decay rate in \nref{decrategr} at leading order in the backreaction. The reason is that they would only redefine $Z_{\text{GR}}(S^{D})$; therefore, any correction factor would cancel between the numerator and denominator of the decay rate formula.

Another point is that with our setup, we could also, in principle, numerically study the backreaction corrections to the decay rate defined via \nref{decrategr}. However, this would require some non-trivial numerical analysis, and we leave this for future work. Also, to do so, one has to argue more carefully for what the volume factor in \nref{decrategr} should be. The reason is that using different volumes that differ from the one we chose by $O(\kappa^{2})$ will lead to a difference of the same order as the corrections to the one-loop determinants of $Z(\text{CdL})$ and $Z_{\fv}(S^{D})$.

\section{Instanton toy model}
\label{toyinstsec}

In this section, we study an instanton toy model where the computation of the one-loop determinant is simpler.  We take the potential $v(\phi)$ to be given by
\begin{equation}
\label{vtoyinst}
v(\phi)=\frac{D(D-2)}{4\kappa^{2}l^{2}}\bigg((1+2\kappa^{2}\epsilon)\cos\bigg(\frac{2\kappa\phi}{\sqrt{(D-2)}}\bigg)-1\bigg)
\end{equation}

And for simplicity, we set $l=1$. One can show that with this choice of potential, there is a simple background profile. Namely, the metric is given by
\begin{equation}
\label{toymodmt}
ds^{2}=\frac{d\rho^{2}}{1+\kappa^{2}\epsilon(1+\sin^{2}\rho)}+\frac{1}{(1+\kappa^{2}\epsilon)}\sin^{2}\rho\, d\Omega_{D-1}^{2}
\end{equation}

We chose not to write it in FRW form because in this form of writing it, the background coordinates $\rho$ range from $0$ to $\pi$ regardless of the parameter $\epsilon$. One can solve for $\phi$ explicitly in terms of $\rho$ as
\begin{equation}
\sin \frac{2\kappa \phi}{\sqrt{D-2}}=-\frac{2\kappa \cos \rho}{1+2\kappa^{2}\epsilon} \sqrt{\epsilon(1+\kappa^{2}\epsilon(1+\sin^{2}\rho))}
\end{equation}

The motivation for studying this solution is that its coupling to the metric in the one-loop determinant can be made quite simple. Namely, by taking the $\gamma=1$ gauge in \nref{gfaction} the only coupling between the scalar fluctuations $\varphi$ and the metric fluctuations $h_{ab}$ is of form $\varphi h^{ab}\n_{a}\n_{b}\phi$. However, in this model it holds that $\n_{a}\n_{b}\phi \sim g_{ab}$, so $\varphi$ only couples to the trace of the metric $h=g^{ab}h_{ab}$. More explicitly, let us decompose the metric fluctuation as
\begin{equation}
h_{ab}=\psi_{ab}+g_{ab}\psi
\end{equation}
with $\psi_{ab}$ traceless. Then, we can check that the pure trace part of the metric $\psi$ does not couple to the traceless part $\psi_{ab}$. Therefore, the eigenvalue equation for $\varphi$ and $\psi$ form a closed system of differential equations. More explicitly, we can find the eigenvalue equation restricted to their components to be
\begin{equation}
\begin{gathered}
\label{toyinst}
\lambda \psi=\frac{(D-2)}{2}\bigg(\n^{2}+2(D-1)+2D \kappa^{2} \epsilon \sin^{2}\rho\bigg)\psi+2 \kappa \cos \rho \sqrt{\epsilon(D-2)(1+\kappa^{2}\epsilon(1+\sin^{2}\rho))}\varphi\\
\lambda \varphi=(-\n^{2}-D-4\kappa^{2}\epsilon \sin^{2}\rho)\varphi+2 \kappa D \cos \rho \sqrt{\epsilon(D-2)(1+\kappa^{2}\epsilon(1+\sin^{2}\rho))}\psi 
\end{gathered}
\end{equation}

\subsection{Background at $\kappa=0$}

Here we discuss the background, and the eigenvalues of the fluctuation operator for $\varphi$ and $\psi$ at $\kappa=0$ and fixed $\epsilon$. At $\kappa=0$ the metric is that of a round sphere $S^{D}$, and the field configuration is
\begin{equation}
\phi_{\kappa=0}=-\cos \rho\sqrt{\epsilon(D-2)}
\end{equation}

We can study \nref{toyinst} at $\kappa=0$ and fixed $\epsilon$. In this limit case, the Laplacian $\n^{2}$ should be replaced by the appropriate round sphere one. The eigenvalues of the sphere Laplacian are $-\ell_{D}(\ell_{D}+D-1)$, with $\ell_{D}$ a non-negative integer standing for the angular momentum of the mode in $S^{D}$. Since at zero backreaction the operator \nref{toyinst} consists only of the sphere Laplacian and constants, the eigenmodes of \nref{toyinst} are, in the notation of \nref{bigphi}, of the form
\begin{equation}
\begin{gathered}
\lambda_{\phi}=\ell_{D}(\ell_{D}+D-1)-D ~~~~,~~~~ \Phi_{\phi}=\begin{pmatrix}
    0\\ f_{\ell_{D}}\\
\end{pmatrix}\\
\lambda_{\psi}=\frac{(D-2)}{2}(-\ell_{D}(\ell_{D}+D-1)+2(D-1)) ~~~~,~~~~~ \Phi_{\psi}=\begin{pmatrix}
    g_{ab} f_{\ell_{D}}\\0
\end{pmatrix}
\end{gathered}
\end{equation}
with $f_{\ell}$ an eigenmode of the sphere Laplacian with angular momentum $\ell_{D}$. 

We note that this system has more zero modes than one would expect. Namely, we expect that the zero modes of \nref{flucgf} should have been only the $D$ modes associated to isometries, but \nref{toyinst} at $\kappa=0$ has an extra zero mode, namely $\varphi=\cos \rho$. This is because the $\kappa=0$ limit of the potential in \nref{vtoyinst} is that of a scalar with negative squared mass, e.g
\begin{equation}
v(\phi)|_{\kappa=0}=\bigg(\frac{D(D-2)}{4\kappa^{2}}\bigg[(1+2\kappa^{2}\epsilon)\bigg(1-\frac{2\kappa^{2}\phi^{2}}{(D-2)}\bigg)-1\bigg]\bigg)_{\kappa=0}=\frac{D}{2}\bigg(\epsilon(D-2)-\phi^{2}\bigg)
\end{equation}

This negative squared mass precisely matches the sphere Laplacian at $\ell_{D}=1$ and thus implies that all $\ell_{D}=1$ modes of $\varphi$ are zero modes in this limit. The modes with $\ell_{D}=1$ along $S^{D}$ divide into modes with $\ell=1$ along $S^{D-1}$, these are the usual zero modes of a generic instanton, and a single mode with $\ell=0$ along $S^{D-1}$, which is the extra zero mode. 

\subsection{Perturbation theory}

In \nref{pertth} we derived a closed formula for the eigenvalue change of the modes associated with shifting $\phi$ by isometries. For $\gamma=1$, and in the context of the potential we are studying, the formula implies\footnote{In principle, the formula would actually have to be corrected since the potential in \nref{vtoyinst} is $\kappa$ dependent. However, the correction is trivial for these isometry modes. The reason is that if there was no backreaction, these modes would be zero modes for any potential, since they follow from the symmetries of $S^{D}$.}
\begin{equation}
\label{predtoy}
\lambda=2\kappa^{2}\bigg(\frac{\int(\xi^{a}\n_{a}\phi)^{2}}{\int \xi_{a}\xi^{a}}\bigg)_{\kappa=0}+O(\kappa^{4})=\kappa^{2}\epsilon(D-2)+O(\kappa^{4})
\end{equation}
where to evaluate the integral, we used some identities from Appendix \nref{useid}.

\begin{figure}[h!]
    \centering
    \includegraphics[width=0.65\linewidth]{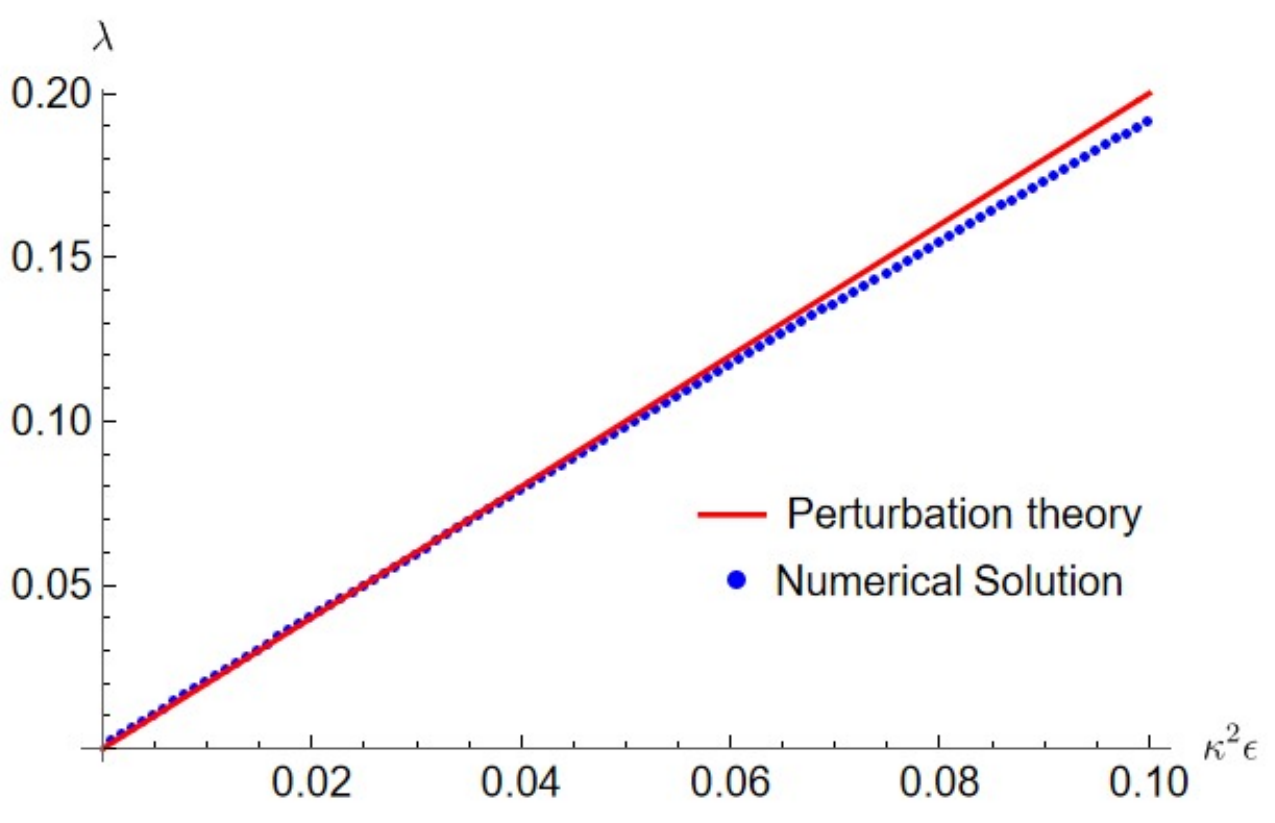}
    \caption{The red line is the eigenvalue shift $\lambda$ computed at leading order in perturbation theory in section \nref{pertth} for $D=4$. The blue points are the numerical results for the eigenvalue we obtained by solving \nref{toyinst} numerically with the \textbf{NDEigenvalues} command. We see that they match precisely up until higher $\kappa$, where the deviation starts becoming relevant.}
    \label{plotpert}
\end{figure}

We can then check this formula by numerically solving equation \nref{toyinst} and comparing it to the prediction \nref{predtoy}. We do so explicitly in Figure \nref{plotpert}, where we used the \textbf{NDEigenvalues} command in Mathematica with \nref{toyinst} as an input to compute the eigenvalues numerically and compare them to \nref{predtoy}. The \textbf{NDEigenvalues} command returns a list of eigenvalues of a differential operator, in increasing distance from zero, so it is a convenient command to plot the lowest lying eigenvalues of a coupled differential operator.

\begin{figure}[h!]
    \centering
    \includegraphics[width=0.65\linewidth]{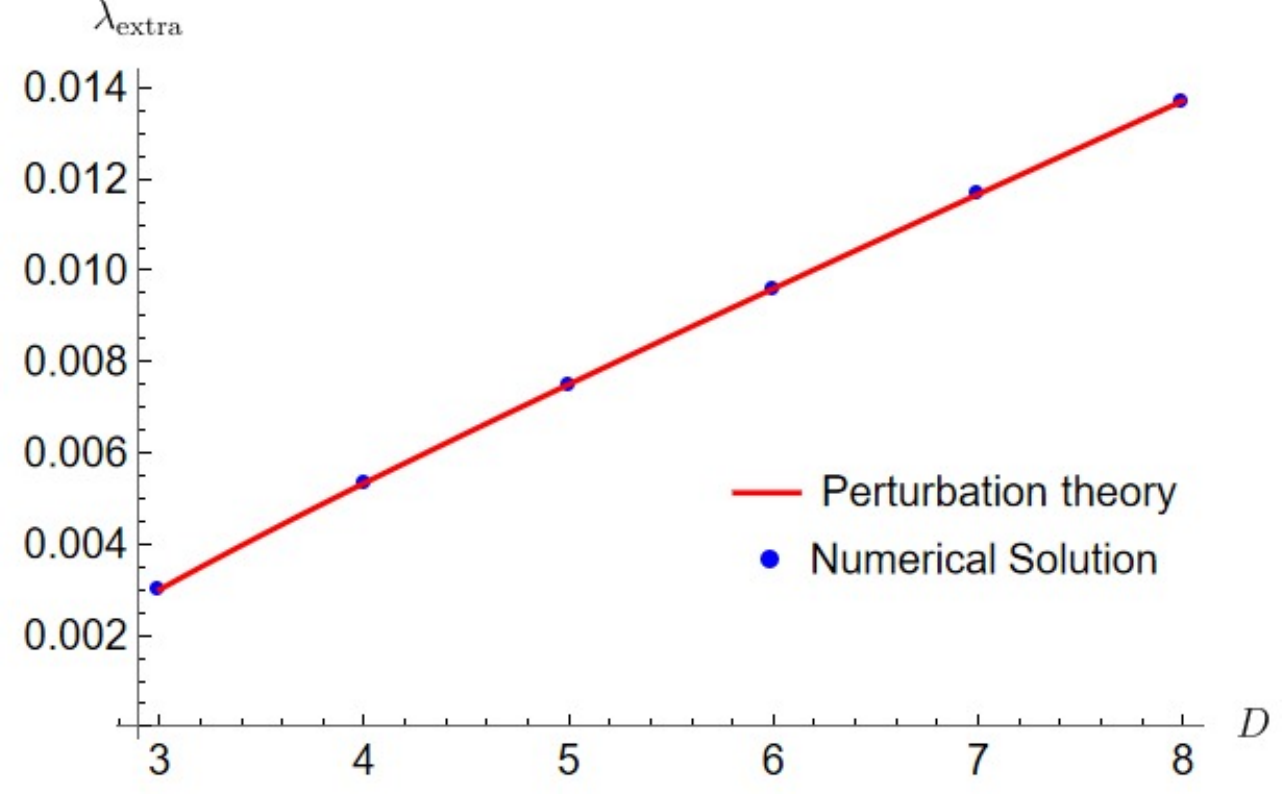}
    \caption{The red line is the eigenvalue shift $\lambda_{\text{extra}}$ of the extra zero mode, computed at leading order in perturbation theory in equation \nref{extralambda}, for $\kappa^{2}\epsilon=0.001$. The blue points are the numerical results for the eigenvalue we obtained by solving \nref{toyinst} numerically with the \textbf{NDEigenvalues} command. We compared the answers for different dimensions $D$.}
    \label{plotpert2}
\end{figure}

We can also discuss how the extra zero mode with $\varphi=\cos \rho$ is lifted. We call its eigenvalue $\lambda_{\text{extra}}$. For this calculation, one has to do the perturbation theory taking into account the fact that the potential $v$ in \nref{vtoyinst} is $\kappa$ dependent. In doing so, one can find that
\begin{equation}
\label{extralambda}
\lambda_{\text{extra}}=\frac{2\kappa^{2}\epsilon D(D-2)}{(D-1)}+O(\kappa^{4})
\end{equation}
where the leading term is actually $\gamma$ independent, as one can check from perturbation theory. This point is important for gauge invariance of the calculation at leading order. We compare this prediction against the numerical evaluation of this eigenvalue in Figure \nref{plotpert2}. Note that $\lambda_{\text{extra}}>0$ for small $\kappa^{2}\epsilon$, so this mode does not lead to an extra phase at small backreaction.

\section{Discussion}
\label{discu}

In the main discussion of the paper, we showed that, once correctly understood, the Euclidean path integral of Coleman de Luccia instantons has the correct features for a saddle associated with tunneling, at least for small enough backreaction. We also show that the decay rate defined from it matches a pure quantum field result at small backreaction. We also discussed how some zero-mode contributions are lifted once a small backreaction breaks the symmetry responsible for them. Now we discuss interesting future directions and other conceptual points.

\subsection{Future directions}

As discussed in the main text, the prescription for dealing with negative modes we used in this paper, also used in \cite{Maldacena:2024spf,Ivo:2025yek}, does not deal correctly with some factors of $2$ that one expects from bounce contributions. Namely, in \cite{Callan:1977pt}, the authors have an extra factor of $\frac{1}{2}$ for the tunneling negative mode integral, as compared to us. This is because of how they define the contour through the bounce saddles. We dealt with this difference by inserting appropriate factors of $\frac{1}{2}$ by hand when necessary.

It would be interesting to understand more generally when negative modes contribute to such factors of $\frac{1}{2}$. For example, if the negative modes in Euclidean gravity also had such factors, it would imply that partition functions, with non-zero phase, computed following \cite{Polchinski:1988ua}, would need corrections. Since this would potentially change the one-loop determinant around some gravity saddles by a finite constant, it seems to be an essential point to understand. We should mention, however, that this point about negative modes in gravity should not affect the decay rates we defined in \nref{gamint}, at least for small backreaction. The reason is that any such correction to these negative mode factors will cancel between the numerator and denominator of the formula.

Another shortcoming is that we have not rigorously addressed is exactly which spacetime volume should appear in the definition of decay rate in \nref{gamint}. The Callan and Coleman argument in \cite{Callan:1977pt} for deriving the decay rate assumes a fixed spacetime geometry, but in gravity, the spacetime volume will be affected by backreaction. In particular, it will be dependent on whether we have a single bounce or multiple bounces in the geometry. The Callan and Coleman argument also seems to assume the bounces to be much smaller than the manifold they are in. Therefore, extending the argument to bounces comparable to the radius of de Sitter also does not seem straightforward.

Another point is that while we evaluated the Euclidean path integral around the Coleman de Luccia saddle, we did not discuss what this saddle is a contribution to. In particular, the connection of the path integral to Lorentzian physics is not manifest, and to use this Euclidean path integral to define decay rates, we extrapolated from what is known in quantum field theory in flat space. It would be interesting to relate the Euclidean path integrals discussed here to vacuum decay defined more fundamentally from real-time physics. 

In particular, in doing so, we should clarify from a more fundamental prescription how we should analytically continue $\hbar$, as in \nref{iepsilonp}, for the purpose of dealing with negative modes. In section \nref{flatsol}, we were able to infer from answer analysis that the correct analytical continuation of $\hbar$ for vacuum decay is
\begin{equation}
\frac{1}{\hbar} \rightarrow \frac{1}{\hbar}(1+i\epsilon)
\end{equation}

However, the correct continuation of $\hbar$ should follow from more fundamental points about the physical observables we are studying. 

\subsection{Phase at finite backreaction}

An interesting question that we have not addressed is how the phase would change as we take the backreaction to become finite and comparable to other scales in the problem.

In a generic coupled matter+gravity system, it is hard to identify from some generic infinite set of negative modes which ones can be "renormalized away", as we did, for example, in section \nref{inftneg}. The reason is that identifying such "renormalizable infinities" involves relating some infinite number of negative modes to a complete set of eigenmodes of a scalar differential operator.

However, for a generic matter+gravity system, when we diagonalize the quadratic action, we are diagonalizing a coupled differential operator acting on both matter and gravity degrees of freedom. This operator will generally have infinitely many eigenvalues of both positive and negative signs, which will not factorize in any obvious way. So, there is a priori no guiding principle on how to "renormalize away" the infinite number of negative modes in this case. 

As we argued in section \nref{pertth}, however, at small backreaction, we can approximate the coupled path integral over the scalar+gravity modes by a product over two factors, and the overall phase is just the sum of the phase for each one. In other words, the reason we could find a phase is that we knew its value at "zero backreaction", where we just had a round sphere and a probe field. Then, we flowed the modes from zero to small, but finite, backreaction. The eigenmodes of the problem at zero backreaction $\Phi_{n}({G_{N}=0})$ flowed smoothly to a set of eigenmodes $\Phi_{n}(G_{N})$ at small backreaction. 
Since no eigenvalue changed from positive to negative, or the converse, we assumed that the phase remained the same. 

Therefore, without the presence of a general guiding principle, we think a reasonable proposal for how to define the phase at finite backreaction is the following "Clutch prescription": We first establish the phase and the eigenmodes of the system at small backreaction, and then we study how they change as we vary the backreaction. More concretely, if we analytically continued Planck's constant as
\begin{equation}
\frac{1}{\hbar}\rightarrow\frac{1}{\hbar}(1\pm i \epsilon)
\end{equation}

It would imply the rules to be as follows: If an eigenvalue changes from positive to negative, the phase gains a factor of $(\pm i)$, and if it changes from negative to positive, the phase gains a factor of $(\mp i)$. In principle, there is a concrete numerical way to find, for example, the smallest value of $\kappa^{2}$ necessary for the phase to change via this rule. 

To see this, note that for the phase to change, it must be that some positive mode becomes negative or the converse happens. By continuity, this can only occur if an eigenvalue crosses zero at some point. So, we can specify how the phase changes by studying how the eigenvalue of the fluctuation operator $M$ closest to the origin, which we call $\lambda_{\text{low}}$, changes as we vary the backreaction.

The phase of the path integral will change only if $\lambda_{\text{low}}$ crosses zero from positive to negative or negative to positive. This can be studied numerically using codes that solve for low-lying eigenvalues of coupled differential operators, such as the \textbf{NDEigenvalues} command on Mathematica used in section \nref{toyinstsec}.

\begin{figure}[h!]
    \centering
    \includegraphics[width=\linewidth]{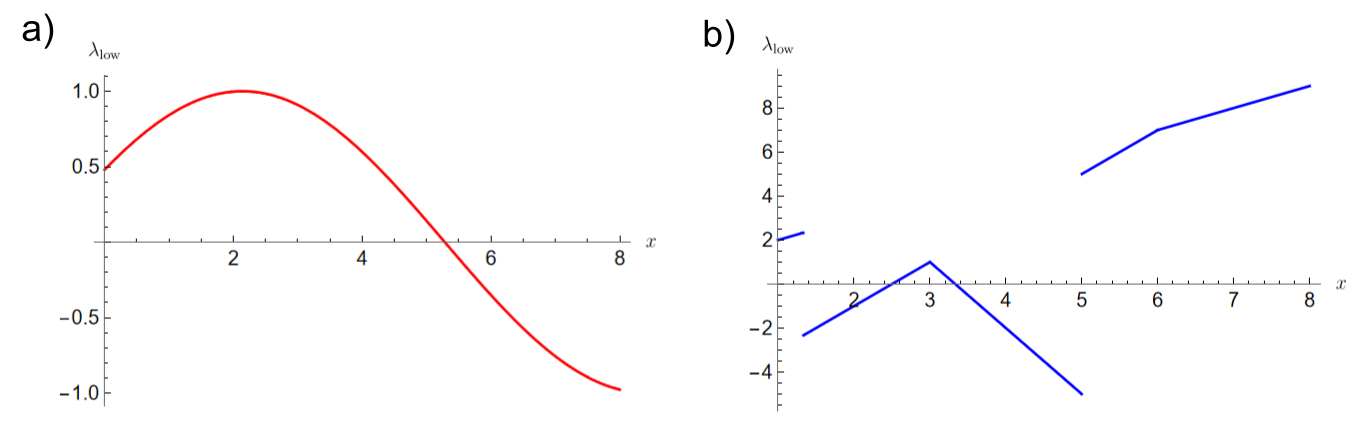}
    \caption{The plots show two possible setups for $\lambda_{\text{low}}$. We imagine that $x$ is some parameter of the theory they were derived from. \textbf{a)} In this plot, $\lambda_{\text{low}}$ is always equal to $\sin(\frac{x+1}{2})$. From $x=0$ to $x=8$, the net result is that $\lambda_{\text{low}}$ crossed from positive to negative once, so the phase changed by $(\pm i)$. \textbf{b)} In this plot, we show a setup where $\lambda_{\text{low}}$ is determined from three competing eigenvalues: $x+1$, $2x-5$ and $10-3x$. We see that $\lambda_{\text{low}}$ has discontinuities, but they do not count as crossings. There were overall zero crossings, so the phase change from $x=0$ to $x=8$ is zero.}
    \label{lambdalow}
\end{figure}

In summary, the Clutch prescription consists of studying how the phase changes by analyzing how the eigenvalue of the fluctuation operator closest to zero behaves. More specifically, by analyzing how many times it crosses zero. Each crossing will change the phase, depending on its orientation. A shortcoming of this prescription is that the phase of zero eigenvalues is sensitive to whether its eigenmodes belong to a mostly positive or a mostly negative field \cite{Ivo:2025yek,Shi:2025amq}. Therefore, to establish the phase change, one should make sure to change the parameters such that the eigenvalues do not end exactly at zero. We display examples of the prescription in Figure \nref{lambdalow}. 

In Figure \nref{lambdalow} we also tried to stress that sometimes $\lambda_{\text{low}}$ will change discontinuously, because which eigenvalue of $M$ is closest to the origin can change as we change the parameters of the theory. However, discontinuities coming from this fact do not count as crossings, since no eigenvalue crossed zero.

This prescription is, of course, not rigorous in any way. But, it is a temporary proposal that seems reasonable for at least a small enough coupling between the matter and metric fluctuations. One would of course like to understand how to define the phase in complicated coupled systems in a more fundamental way. Such a procedure might not even exist, but any positive or negative results in this direction would be interesting.

\subsection{Adding observers?}

A possibly interesting question is how, for example, one could add semiclassical observers to bubble-nucleating spacetimes, such as the ones we are discussing. More specifically, what is the equivalent of the observer discussion of \cite{Maldacena:2024spf}? The spacetimes we are studying are FRWs, so their line element is of the form
\begin{equation}
ds^{2}=d\chi^{2}+a^{2}(\chi)d\Omega_{D-1}^{2}
\end{equation}
with $d\Omega_{D-1}^{2}$ the line element in $S^{D-1}$. We will refer to a FRW geometry that follows from a Coleman de Luccia solution as a "Coleman de Luccia manifold".

The semiclassical observer must be along a geodesic, so the first thing to discuss is what the geodesics are in this spacetime. The simplest non-trivial geodesic we can construct, which wraps once as the one in \cite{Maldacena:2024spf}, is a great circle in $S^{D-1}$. For it to be a geodesic in the FRW, we also need to pick $\chi=\bar{\chi}$ such that $a(\bar{\chi})$ is a local maximum of $a$. Perhaps a more convenient way of writing the line element is as follows
\begin{equation}
ds^{2}=d\chi^{2}+a^{2}(\chi)(\cos^{2}\theta\,d\alpha^{2}+d\theta^{2}+\sin^{2}\theta\,d\Omega_{D-3}^{2})
\end{equation}

\begin{figure}[h!]
    \centering
    \includegraphics[width=0.8\linewidth]{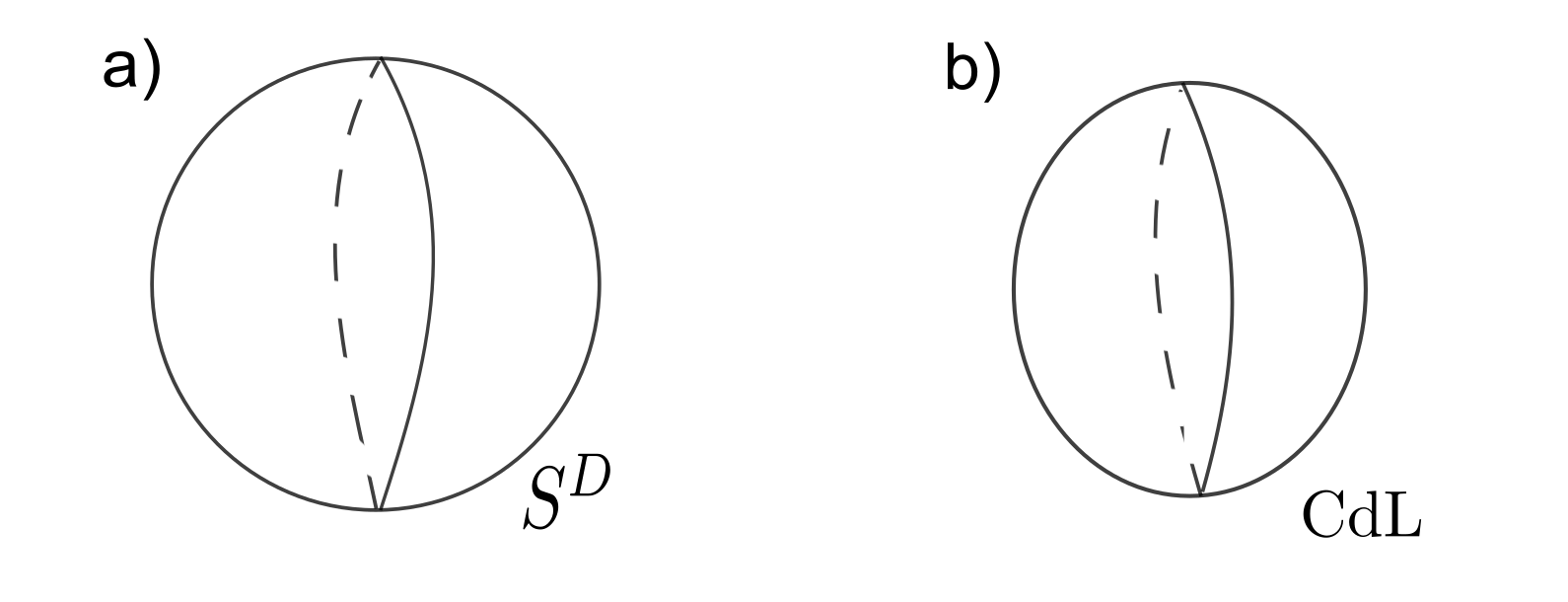}
    \caption{\textbf{a)} The round sphere $S^{D}$ with an observer. This is the type of Euclidean solution one would use to prepare states for observers in de Sitter. \textbf{b)} Coleman de Luccia manifold with an observer. The geometry is deformed with respect to its round sphere counterpart because of the scalar backreaction; in particular, it is more concave at the top. Note also that since $\phi$ depends only on $\chi$, the value of the scalar along the observer's trajectory is constant.}
    \label{obsbub}
\end{figure}

In terms of this decomposition, we pick the geodesic defined by $\theta=0$ and $\chi=\bar{\chi}$, and extended along the $\alpha$ direction. $\alpha$ goes from $0$ to $2\pi$. The geodesic is displayed in figure \nref{obsbub}, where we also compare it to its round sphere counterpart.

Another interesting point to discuss is how one can construct states, including observers, from this saddle with a non-trivial geodesic, as was done in \cite{Chandrasekaran:2022cip,Witten:2023xze}. The idea is to cut the manifold into a plane of symmetry $W$, and then continue the solution to Lorentzian signature, viewing $W$ as an initial value surface. We see that the geodesic intersects $W$ in two different points, which implies that the state contains two observers. The line element in Lorentzian signature is
\begin{equation}
ds^{2}=d\chi^{2}+a^{2}(\chi)(-\cos^{2}\theta\,dt^{2}+d\theta^{2}+\sin^{2}\theta\, d\Omega_{D-3}^{2})
\end{equation}

We can read this metric as the static patch metric of a lower-dimensional de Sitter, and an "extra direction" $d\chi^{2}$.  The geodesic continues to Lorentzian signature from $\alpha=0$ and $\alpha=\pi$ by taking $\alpha=\pi+i t_{R}$ to parameterize the observer in the right, and $\alpha=-i t_{L}$ to parameterize the observer in the left. 

\subsubsection{Negative modes}

An interesting point is that this observer treated semiclassically will contribute to the Euclidean action as $e^{-mL}$, with $L$ the length of the geodesic. Therefore, if fluctuations of $L$ have negative modes, the observer will contribute to the phase accordingly, as in \cite{Maldacena:2024spf}. 

By deforming the observer's trajectory, while using the coordinate $\alpha$ to parameterize the curve, we can find the deformation of the length, $\delta L$, to be
\begin{equation}
\delta L=\frac{1}{2}\int_{0}^{2\pi}d\alpha \bigg[\frac{1}{a(\bar{\chi})}\delta \chi(-\partial_{\alpha}^{2}+a(\bar{\chi})\ddot{a}(\bar{\chi}))\delta \chi+a(\bar{\chi})\sum_{i=1}^{D-2}\delta \theta_{i}(-\partial_{\alpha}^{2}-1)\delta \theta_{i}\bigg]
\end{equation}
where dot stands for $\dot{f}=\partial_{\chi}f$, and $\delta \theta_{i}$ is a component of the vector $\delta\vec{\theta}$ that correspond to fluctuations of the observer along $S^{D-2}$. This is the same $\vec{\theta}$ notation used in \cite{Maldacena:2024spf}, except here $\delta \vec{\theta}$ is defined in one lower dimension. 

Decomposing the fluctuations into a winding mode basis $e^{\pm i n \alpha}$, with $n$ an integer, we find that there are $D-2$ negative modes coming from the $\delta \theta_{i}$ modes, more specifically from the modes with $n=0$. Also, their $n=\pm 1$ modes are zero modes.

Then, we need to discuss the negative modes of $\delta \chi$. For this, one needs to examine its "mass term", $a(\bar{\chi})\ddot{a}(\bar{\chi})$, first. One can fix this term using Friedman's equations. In doing so, we can show that
\begin{equation}
a(\bar{\chi})\ddot{a}(\bar{\chi})=-\frac{(1+(D-2)\bar{\epsilon})}{1-\bar{\epsilon}}<-1, \text{with}~~~~~~ \bar{\epsilon}=\frac{\dot{\phi}^{2}(\bar{\chi})}{\bigg(2v(\phi(\bar{\chi}))+\frac{(D-1)(D-2)}{\kappa^{2}}\bigg)}
\end{equation}
with $\bar{\epsilon}$ a slow-roll parameter of the solution evaluated at $\chi=\bar{\chi}$. The term in the denominator of $\bar{\epsilon}$ comes from the scalar potential $v(\phi)$ and an offset of the potential defined by the cosmological constant $\Lambda$. The $\Lambda$ was chosen such that the geometry is a sphere with unit radius as $\kappa \rightarrow 0$.

Therefore, since $\bar{\epsilon} \geq 0$, as long as $\bar{\epsilon}$ is not zero, the mass term $a(\bar{\chi})\ddot{a}(\bar{\chi})$ will be less than $-1$. This implies that in some sense, the Coleman de Luccia manifold is always more concave at its waist than the round sphere. 

If the backreaction is small, $a(\bar{\chi})\ddot{a}(\bar{\chi})$ will be very close to $-1$.  Therefore, we conclude that, at small backreaction, the $\delta \chi$ fluctuations have three negative modes, one coming from $n=0$ and two others coming from $n=\pm 1$. These $n=\pm 1$ modes would be zero modes if $\dot{\phi}(\bar{\chi})$ were zero. Therefore, the rolling scalar field in the background induces the observer geodesic in the Coleman de Luccia manifold to have two more negative modes than the one in \cite{Maldacena:2024spf}. 

Having discussed these points, we can also discuss the phase of this saddle’s contribution, $\rho$, to the observer’s density of states, in the small backreaction limit. Namely, if we follow the construction of \cite{Maldacena:2024spf} to define the density of states, we first have to go to the microcanonical ensemble, which gives a factor of $(\pm i)$. Then, the rest of the phase comes from the path integral over the relevant saddle and the contribution from the observer's geodesic. This would then allow us to conclude that for $\rho$
\begin{equation}
\label{densitma}
\rho \approx (\pm i)(\pm i)(\mp i)^{D+2}(\pm i)^{D-1+2}(\text{positive})=(\pm i)(\text{positive})
\end{equation}
where the first phase factor is microcanonical, the second is from the bubble negative mode in $Z_{\phi}(\text{bounce})$, the third is from the sphere path integral $Z_{\text{GR}}(S^{D})$, and the last is from the observer. Interestingly enough, this contribution to the density of states has the same phase as the path integral over the bounce.  

Another way we can see that is by noting that an observer in de Sitter would have a phase of $(-1)$ according to \cite{Maldacena:2024spf}. This minus one cancels with the extra phase $(\pm i)^{2}$ from the $\chi$ modes, induced by the backreaction. Then, the only remaining phase comes from the negative mode in $Z_{\phi}(\text{bounce})$. It would be interesting to understand if this type of extra phase coming from observer modes could be leveraged as a way of canceling the phase of $(-1)$ found in \cite{Maldacena:2024spf} for an observer in de Sitter\footnote{The minus sign discussed here follows from the original prescription of \cite{Maldacena:2024spf}. An alternative prescription, which gives a positive density of states, was later proposed in \cite{Chen:2025jqm}. See the discussion of \cite{Ivo:2025xek} for further comments.}.

\subsection*{Acknowledgments}

We would like to specially thank Juan Maldacena for helpful discussions and very valuable advice on the manuscript. We would also like to thank Tom Banks, Yiming Chen, Zimo Sun and Edward Witten for discussions. We would also like to thank Yiming Chen and Zimo Sun for comments on the manuscript. During revision for submission to JHEP, ChatGPT was used to assist with proofreading, literature checks, and consistency checks. The author takes responsibility for the final manuscript.

\appendix

\section{Gauge fixing normalization}
\label{gfnorm}

Here, we briefly discuss the normalization of the gauge-fixed path integral when gauge-fixing terms are added. First, note that at the heuristic level, the path integral over metrics and a set of matter fields $\phi^{I}$ can be defined as  
\begin{equation}
\label{zdefap}
Z=\frac{\int Dg D\phi^{I}e^{-I_{E}}}{\int D\xi^{a}}
\end{equation}
where $\int D\xi^{a}$ stands for the path integral over the diffeomorphism group. Still at the level of heuristics, the ghost determinant for gravity comes from inserting in the path integral some identity of the form of
\begin{equation}
\label{ghid}
\int D'\xi^{a}\, \text{Det'}(\delta_{\xi^a}P^{b}) \delta'(P^{b}-\omega^{b})=1
\end{equation}
with $P_{b}-\omega_{b}=0$ a gauge condition on the fields, $a$ and $b$ some spacetime indices, and $\delta_{\xi}P^{a}$ the functional derivative of $P$ along the gauge transformation induced by $\xi$. 

Also, note that we used $D'\xi^{a}$ to mean we do the path integral over diffeomorphisms with isometries of the background excluded from the path integral. This is because we wish to deal with them differently. The prime notation in other quantities means the same. That is, $\text{Det}'$ stands for a functional determinant with isometry contributions omitted, and the same for the delta function $\delta'$. For consistency, we also assume that both $P^{a}$ and $\omega^{a}$ themselves are orthogonal to the isometries, in an appropriate local norm. 

We should also mention that for the integral over group volume in \nref{ghid} to match with the one from the path integral in \nref{zdefap} mode by mode, we should pick the path integral measure for diffeomorphisms in \nref{ghid} to be the same as the one in \nref{zdefap}. This then also fixes the normalization of the ghost determinant, $\text{Det}'(\delta_{\xi}P^{a})$. We define the measure over diffeomorphisms $D\xi^{a}$ by taking it to follow from the local norm
\begin{equation}
(\xi^{a},\xi^{a})=\int \xi^{a}\xi_{a}
\end{equation}
and by requiring it to be normalized such that
\begin{equation}
\label{normmeas}
\int D\xi^{a} e^{-\frac{1}{2}(\xi^{a},\xi^{a})}=1
\end{equation}

Alternatively, if one can decompose $\xi^{a}$ in a complete basis of functions $\xi=\sum_{n}c_{n}\xi^{a}_{n}$, with $\xi_{n}^{a}$ unit normalized on the manifold, then
\begin{equation}
D\xi^{a}=\prod_{n} \frac{dc_{n}}{\sqrt{2\pi}}
\end{equation}

Therefore, for the delta function to be consistent, one needs it to be expressed as
\begin{equation}
\delta'(P^{a}-\omega^{a})=\int \prod_{n}'\frac{dB_{n}}{\sqrt{2\pi}} e^{iB_{n}(P_{n}-\omega_{n})}=\int D'B^{a} e^{i(B^{a},P_{a}-\omega_{a})}
\end{equation}
where we decomposed $P^{a}$ and $\omega^{a}$ on the same basis and we introduced the auxiliary field $B=\sum_{n}B_{n}\xi_{n}^{a}$. The measure for $B$ was fixed by the delta function condition against the measure $D\omega^{a}$ defined by \nref{normmeas}, and we used the prime notation again to mean that isometries are excluded.

This fixes the discussion of the determinant. Since \nref{ghid} is true for any $\omega$, it should be true for an average over $\omega^{a}$ as well, which one can insert by including an integral over $\omega^{a}$ as
\begin{equation}
N(\sigma)\int D'\omega \, e^{-\frac{1}{2\sigma^{2}}(\omega,\omega)}=1
\end{equation}
where the measure $D\omega$ is defined as in \nref{normmeas}. This means that $N(\sigma)=\prod_{n}' \sigma^{-1}$. Since the infinite product in $N$ has a few modes missing, it is not a local contribution. Therefore, it has an overall finite part, which we have to take into account. This was discussed, for example, in \cite{Donnelly:2013tia}, where it was argued that we need to keep these overall factors from $N(\sigma)$ in the path integral for consistency.

In the case of this paper, however, it seems we have a more convenient option. The modes that appear in the $P^{a}$ and the $\xi^{a}$ integrals are all the modes in the mode expansion of vectors in the manifold, except for the isometries. Since $N(\sigma)$ contains a factor of $\sigma^{-1}$ for each of these modes, we can absorb $N(\sigma)$ in the ghost determinant by scaling its argument as $P \rightarrow \sigma^{-1}P$, and therefore
\begin{equation}
Z=\frac{\int Dg D\phi^{I}e^{-I_{E}}}{\int D\xi^{a}}=Z_{0}\int Dg D\phi^{I}\, \text{Det}'(\sigma^{-1}\delta_{\xi}P^{a}) e^{-I_{E}-\frac{1}{2\sigma^{2}}(P^{a},P^{a})}
\end{equation}
which fixes a convenient normalization for the ghost determinant for a given gauge fixing term added to the action. The factor $Z_{0}$ comes from the path integral over isometries in the denominator.

It should be intuitively clear why the factor of $\sigma^{-1}$ in the ghost determinant is natural. That is, we can write the following correct identity
\begin{equation}
1=\int D'y e^{-\frac{1}{2}(y,y)}=\int D'(\sigma^{-1}P) e^{-\frac{1}{2\sigma^{2}}(P,P)}=\int D'\xi^{a}\,\text{Det}'(\sigma^{-1}\delta_{\xi}P^{a})e^{-\frac{1}{2\sigma^{2}}(P^{a},P^{a})}
\end{equation}
where the first identity is true by construction for a vector field $y$ with this measure. Then, we did the field redefinition $y=\sigma^{-1}P$, and we rewrote the path integral appropriately. 

Note that in omitting the isometries we have to study their contribution separately, e.g, in the zero mode path integral $Z_{0}$. In the main text, the relevant zero mode path integral was for the Coleman de Luccia instanton, where the zero modes are the $SO(D)$ generators.

To find the path integral over these zero modes, we follow the strategy of \cite{Anninos:2020hfj,Law:2020cpj} where we relate the path integral measure to the canonical group measure in this group. The first point is to note that because of the rescaling we did in $h_{ab}$ and how we defined the gauge transformations in the main text \nref{gaugetr}, the group Lie algebra $[[,]]$ induced by the path integral is
\begin{equation}
\label{pialg}
[[v,u]]=\sqrt{2}\kappa[v,u]
\end{equation}

That is, the bracket in the left hand side stands for the group Lie algebra following from the path integral, and the bracket in the right hand side stands for the spacetime bracket $[v,u]_{a}=v^{c}\n_{c}u_{a}-u^{c}\n_{c}v_{a}$. Therefore, to pick generators whose path integral induced algebra \nref{pialg} satisfy the commutation relations of $SO(D)$, we first pick $\xi^{a}$ that satisfy the $SO(D)$ commutation relations in the spacetime algebra. Then, we rescale it by $\frac{1}{\sqrt{2}\kappa}$.

The generators that satisfy the commutation relations of $SO(D)$ are unit normalized in the canonical group norm, $ds_{c}^{2}$, but are not unit normalized in the path integral norm. So, to relate the two, we evaluate the norm of these generators as
\begin{equation}
ds^{2}=\bigg(\frac{\xi^{a}}{\sqrt{2}\kappa},\frac{\xi^{a}}{\sqrt{2}\kappa}\bigg)=\frac{1}{2\kappa^{2}}(\xi^{a},\xi^{a})ds_{c}^{2}
\end{equation}
where we used that the generators are unit normalized in the canonical norm $ds_{c}^{2}$. Therefore, the relation between the path integral over isometries and the canonical group volume is
\begin{equation}
\text{Vol}(G)_{\text{PI}}=\bigg(\frac{(\xi^{a},\xi^{a})}{2\pi(\sqrt{2}\kappa)^{2}}\bigg)^{\frac{|SO(D)|}{2}}\text{Vol} (SO(D))_{c}
\end{equation}
with $|SO(D)|$ the dimension of $SO(D)$. The contribution of the ghost zero modes is the same as the inverse of the integral over the isometries in the path integral measure; therefore
\begin{equation}
Z_{0}=\frac{1}{\text{Vol}(G)_{PI}}=\frac{1}{\text{Vol}(SO(D))_{c}}\bigg(\frac{2\pi (\sqrt{2}\kappa)^{2}}{(\xi^{a},\xi^{a})}\bigg)^{\frac{|SO(D)|}{2}}
\end{equation}

\section{Useful identities}
\label{useid}

Here we discuss some useful identities about the isometries in the sphere $S^{D}$, which were discussed in the main text. As a reminder, we take the line element to be
\begin{equation}
ds^{2}=d\tau^{2}+\sin^{2}\tau\, d\Omega_{D-1}^{2}
\end{equation}

We consider the isometries that have some $\tau$ component; they are of the form
\begin{equation}
\xi_{a}=Y_{1}\, n_{a}+\sin\tau\, \cos \tau\, \n_{a}Y_{1}
\end{equation}
with $n^{a}=(\partial_{\tau})^{a}$. The $Y_{1}$ is a mode with $\ell=1$ angular momentum along the $S^{D-1}$ and normalized such that they match one of the embedding coordinates of $S^{D-1}$. Namely we can write $S^{D-1}$ as the surface in $R^{D}$ that satisfy
\begin{equation}
\sum_{i=1}X_{i}^{2}=1
\end{equation}
with $X_{i}$ the Euclidean coordinates in $R^{D}$. There are therefore $D$ linearly independent $\ell=1$ modes of the form we discussed, one for each coordinate $X^{i}$. 

It will be useful for us to evaluate the local norm of $\xi^{a}$
\begin{equation}
(\xi^{a},\xi^{a})=\int \xi^{a}\xi_{a}=\int d\tau\,d\Omega_{D-1}\, \sin ^{D-1}\tau\, (Y_{1}^{2}+\sin^{2}\tau \, \cos^{2}\tau\,\n_{b}Y_{1}\,\n^{b}Y^{1})
\end{equation}

Since $Y_{1}$ is dependent only on $S^{D-1}$ we can replace the covariant derivatives $\n_{a}$ with the tangent covariant derivatives $D_{a}$ induced in the $S^{D-1}$ hypersurfaces. Then, using that $Y_{1}$ is an $l=1$ mode it holds that
\begin{equation}
D^{a}D_{a}Y_{1}=-\frac{\ell(\ell+D-2)}{\sin^{2} \tau}Y_{1}=-\frac{(D-1)}{a^{2}}Y_{1}
\end{equation}
where one can think of the $\sin^{2}\tau$ in the denominator as coming from the local size of the $S^{D-1}$. Using this and integrating by parts, we can write
\begin{equation}
(\xi^{a},\xi^{a})=\int \xi^{a}\xi_{a}=\int d\tau\,d\Omega_{D-1}\, \sin ^{D-1}\tau\, (1+(D-1)\cos^{2}\tau)Y_{1}^{2}
\end{equation}

Using integration by parts twice, we can show that
\begin{equation}
\int_{0}^{\pi} d\tau\sin^{D-1}\tau\, \cos^{2}\tau=\frac{1}{(D+1)}\int_{0}^{\pi} d\tau\sin^{D-1}\tau
\end{equation}

Also, since $Y_{1}$ is an embedding coordinate in $R^{D}$, it must be that the sum over squares of $D$ of these coordinates is one. Therefore, the average over each squared embedding coordinate along $S^{D-1}$ must be $\frac{1}{D}$. Using this fact, we then find that
\begin{equation}
\label{xinorm}
(\xi^{a},\xi^{a})=\int \xi^{a}\xi_{a}=\frac{2D}{(D+1)}\frac{1}{D}\int d\tau\,d\Omega_{D-1}\, \sin ^{D-1}\tau\, =\frac{2}{(D+1)}\text{Vol}(S^{D})
\end{equation}
where we recognize the last integral as the sphere volume. Another way one can obtain this result is that one can write $\xi^{a}$ in terms of embedding coordinates in $R^{D+1}$, where they are of the form $X_{i}\partial_{j}-X_{j}\partial_{i}$. Their squared norm is therefore the integral of $X_{i}^{2}+X_{j}^{2}$. Each term has average of $\frac{1}{(D+1)}$ along $S^{D}$, and therefore \nref{xinorm} follows.

\bibliographystyle{apsrev4-1long}
\bibliography{main.bib}

@article{Coleman:1977py,
    author = "Coleman, Sidney R.",
    title = "{The Fate of the False Vacuum. 1. Semiclassical Theory}",
    reportNumber = "HUTP-77-A004",
    doi = "10.1103/PhysRevD.16.1248",
    journal = "Phys. Rev. D",
    volume = "15",
    pages = "2929--2936",
    year = "1977",
    note = "[Erratum: Phys.Rev.D 16, 1248 (1977)]"
}

@article{Callan:1977pt,
    author = "Callan, Jr., Curtis G. and Coleman, Sidney R.",
    title = "{The Fate of the False Vacuum. 2. First Quantum Corrections}",
    reportNumber = "HUTP-77-A032",
    doi = "10.1103/PhysRevD.16.1762",
    journal = "Phys. Rev. D",
    volume = "16",
    pages = "1762--1768",
    year = "1977"
}

@article{Coleman:1980aw,
    author = "Coleman, Sidney R. and De Luccia, Frank",
    title = "{Gravitational Effects on and of Vacuum Decay}",
    reportNumber = "SLAC-PUB-2463",
    doi = "10.1103/PhysRevD.21.3305",
    journal = "Phys. Rev. D",
    volume = "21",
    pages = "3305",
    year = "1980"
}

@article{Hawking:1981fz,
    author = "Hawking, S. W. and Moss, I. G.",
    title = "{Supercooled Phase Transitions in the Very Early Universe}",
    reportNumber = "Print-82-0181 (CAMBRIDGE)",
    doi = "10.1016/0370-2693(82)90946-7",
    journal = "Phys. Lett. B",
    volume = "110",
    pages = "35--38",
    year = "1982"
}

@article{Jensen:1983ac,
    author = "Jensen, Lars Gerhard and Steinhardt, Paul Joseph",
    title = "{Bubble Nucleation and the {Coleman-Weinberg} Model}",
    reportNumber = "UPR-0231T",
    doi = "10.1016/0550-3213(84)90021-X",
    journal = "Nucl. Phys. B",
    volume = "237",
    pages = "176--188",
    year = "1984"
}

@article{Jensen:1988zx,
    author = "Jensen, Lars Gerhard and Steinhardt, Paul Joseph",
    title = "{Bubble Nucleation for Flat Potential Barriers}",
    reportNumber = "CERN-TH-5183/88",
    doi = "10.1016/0550-3213(89)90539-7",
    journal = "Nucl. Phys. B",
    volume = "317",
    pages = "693--705",
    year = "1989"
}

@article{Coleman:1987rm,
    author = "Coleman, Sidney R.",
    title = "{Quantum Tunneling and Negative Eigenvalues}",
    reportNumber = "HUTP-87/A059",
    doi = "10.1016/0550-3213(88)90308-2",
    journal = "Nucl. Phys. B",
    volume = "298",
    pages = "178--186",
    year = "1988"
}

@article{Lavrelashvili:1985vn,
    author = "Lavrelashvili, George V. and Rubakov, V. A. and Tinyakov, P. G.",
    title = "{TUNNELING TRANSITIONS WITH GRAVITATION: BREAKING OF THE QUASICLASSICAL APPROXIMATION}",
    doi = "10.1016/0370-2693(85)90761-0",
    journal = "Phys. Lett. B",
    volume = "161",
    pages = "280--284",
    year = "1985"
}

@article{Tanaka:1992zw,
    author = "Tanaka, Takahiro and Sasaki, Misao",
    title = "{False vacuum decay with gravity: Negative mode problem}",
    reportNumber = "KUNS-1135",
    doi = "10.1143/PTP.88.503",
    journal = "Prog. Theor. Phys.",
    volume = "88",
    pages = "503--528",
    year = "1992"
}

@article{Garriga:1993fh,
    author = "Garriga, Jaume",
    title = "{Nucleation rates in flat and curved space}",
    eprint = "hep-ph/9308280",
    archivePrefix = "arXiv",
    reportNumber = "TUTP-10-93",
    doi = "10.1103/PhysRevD.49.6327",
    journal = "Phys. Rev. D",
    volume = "49",
    pages = "6327--6342",
    year = "1994"
}

@article{Tanaka:1998mp,
    author = "Tanaka, Takahiro and Sasaki, Misao",
    title = "{No supercritical supercurvature mode conjecture in one bubble open inflation}",
    eprint = "gr-qc/9808018",
    archivePrefix = "arXiv",
    reportNumber = "OU-TAP-82",
    doi = "10.1103/PhysRevD.59.023506",
    journal = "Phys. Rev. D",
    volume = "59",
    pages = "023506",
    year = "1999"
}

@article{Tanaka:1999pj,
    author = "Tanaka, Takahiro",
    title = "{The No - negative mode theorem in false vacuum decay with gravity}",
    eprint = "gr-qc/9901082",
    archivePrefix = "arXiv",
    reportNumber = "OU-TAP-92",
    doi = "10.1016/S0550-3213(99)00369-7",
    journal = "Nucl. Phys. B",
    volume = "556",
    pages = "373--396",
    year = "1999"
}

@article{Lavrelashvili:1999sr,
    author = "Lavrelashvili, George V.",
    editor = "de Alfaro, Vittorio and Nelson, J. E. and Cadoni, M. and Cavaglia, M. and Filippov, A.",
    title = "{Negative mode problem in false vacuum decay with gravity}",
    eprint = "gr-qc/0004025",
    archivePrefix = "arXiv",
    doi = "10.1016/S0920-5632(00)00756-8",
    journal = "Nucl. Phys. B Proc. Suppl.",
    volume = "88",
    pages = "75--82",
    year = "2000"
}

@article{Khvedelidze:2000cp,
    author = "Khvedelidze, Arsen and Lavrelashvili, George V. and Tanaka, Takahiro",
    title = "{On cosmological perturbations in closed FRW model with scalar field and false vacuum decay}",
    eprint = "gr-qc/0001041",
    archivePrefix = "arXiv",
    doi = "10.1103/PhysRevD.62.083501",
    journal = "Phys. Rev. D",
    volume = "62",
    pages = "083501",
    year = "2000"
}

@article{Gratton:2000fj,
    author = "Gratton, Steven and Turok, Neil",
    title = "{Homogeneous modes of cosmological instantons}",
    eprint = "hep-th/0008235",
    archivePrefix = "arXiv",
    doi = "10.1103/PhysRevD.63.123514",
    journal = "Phys. Rev. D",
    volume = "63",
    pages = "123514",
    year = "2001"
}

@article{Dunne:2006bt,
    author = "Dunne, Gerald V. and Wang, Qing-hai",
    title = "{Fluctuations about Cosmological Instantons}",
    eprint = "hep-th/0605176",
    archivePrefix = "arXiv",
    doi = "10.1103/PhysRevD.74.024018",
    journal = "Phys. Rev. D",
    volume = "74",
    pages = "024018",
    year = "2006"
}

@article{Lee:2014uza,
    author = "Lee, Hakjoon and Weinberg, Erick J.",
    title = "{Negative modes of Coleman-De Luccia bounces}",
    eprint = "1408.6547",
    archivePrefix = "arXiv",
    primaryClass = "hep-th",
    doi = "10.1103/PhysRevD.90.124002",
    journal = "Phys. Rev. D",
    volume = "90",
    number = "12",
    pages = "124002",
    year = "2014"
}

@article{Koehn:2015hga,
    author = "Koehn, Michael and Lavrelashvili, George and Lehners, Jean-Luc",
    title = "{Towards a Solution of the Negative Mode Problem in Quantum Tunnelling with Gravity}",
    eprint = "1504.04334",
    archivePrefix = "arXiv",
    primaryClass = "hep-th",
    doi = "10.1103/PhysRevD.92.023506",
    journal = "Phys. Rev. D",
    volume = "92",
    number = "2",
    pages = "023506",
    year = "2015"
}

@article{Banks:2002nm,
    author = "Banks, T.",
    title = "{Heretics of the false vacuum: Gravitational effects on and of vacuum decay. 2.}",
    eprint = "hep-th/0211160",
    archivePrefix = "arXiv",
    reportNumber = "RUNHETC-2002-44, SCIPP-02-27",
    month = "11",
    year = "2002"
}

@article{Hackworth:2004xb,
    author = "Hackworth, James C. and Weinberg, Erick J.",
    title = "{Oscillating bounce solutions and vacuum tunneling in de Sitter spacetime}",
    eprint = "hep-th/0410142",
    archivePrefix = "arXiv",
    reportNumber = "CU-TP-1118",
    doi = "10.1103/PhysRevD.71.044014",
    journal = "Phys. Rev. D",
    volume = "71",
    pages = "044014",
    year = "2005"
}

@article{Gibbons:1976ue,
    author = "Gibbons, G. W. and Hawking, S. W.",
    title = "{Action Integrals and Partition Functions in Quantum Gravity}",
    reportNumber = "PRINT-76-0995 (CAMBRIDGE)",
    doi = "10.1103/PhysRevD.15.2752",
    journal = "Phys. Rev. D",
    volume = "15",
    pages = "2752--2756",
    year = "1977"
}

@article{Gibbons:1978ac,
    author = "Gibbons, G. W. and Hawking, S. W. and Perry, M. J.",
    title = "{Path Integrals and the Indefiniteness of the Gravitational Action}",
    reportNumber = "PRINT-78-0375 (CAMBRIDGE)",
    doi = "10.1016/0550-3213(78)90161-X",
    journal = "Nucl. Phys. B",
    volume = "138",
    pages = "141--150",
    year = "1978"
}

@book{Hawking:2010nzr,
    editor = "Hawking, Stephen W. and Israel, E.",
    title = "{General Relativity}: {An Einstein Centenary Survey, Pt.2}",
    isbn = "978-0-521-13801-7, 978-0-521-13798-0",
    publisher = "Univ. Pr.",
    address = "Cambridge, UK",
    year = "2010"
}

@article{Polchinski:1988ua,
    author = "Polchinski, Joseph",
    title = "{The phase of the sum over spheres}",
    reportNumber = "UTTG-31-88",
    doi = "10.1016/0370-2693(89)90387-0",
    journal = "Phys. Lett. B",
    volume = "219",
    pages = "251--257",
    year = "1989"
}

@article{Maldacena:2024spf,
    author = "Maldacena, Juan",
    title = "{Real observers solving imaginary problems}",
    eprint = "2412.14014",
    archivePrefix = "arXiv",
    primaryClass = "hep-th",
    month = "12",
    year = "2024"
}

@article{Ivo:2025yek,
    author = "Ivo, Victor and Maldacena, Juan and Sun, Zimo",
    title = "{Physical instabilities and the phase of the Euclidean path integral}",
    eprint = "2504.00920",
    archivePrefix = "arXiv",
    primaryClass = "hep-th",
    month = "4",
    year = "2025"
}

@article{Shi:2025amq,
    author = "Shi, Xiaoyi and Turiaci, Gustavo J.",
    title = "{The phase of the gravitational path integral}",
    eprint = "2504.00900",
    archivePrefix = "arXiv",
    primaryClass = "hep-th",
    doi = "10.1007/JHEP07(2025)047",
    journal = "JHEP",
    volume = "07",
    pages = "047",
    year = "2025"
}

@article{Donnelly:2013tia,
    author = "Donnelly, William and Wall, Aron C.",
    title = "{Unitarity of Maxwell theory on curved spacetimes in the covariant formalism}",
    eprint = "1303.1885",
    archivePrefix = "arXiv",
    primaryClass = "hep-th",
    doi = "10.1103/PhysRevD.87.125033",
    journal = "Phys. Rev. D",
    volume = "87",
    number = "12",
    pages = "125033",
    year = "2013"
}

@article{Law:2020cpj,
    author = "Law, Y. T. Albert",
    title = "{A compendium of sphere path integrals}",
    eprint = "2012.06345",
    archivePrefix = "arXiv",
    primaryClass = "hep-th",
    doi = "10.1007/JHEP12(2021)213",
    journal = "JHEP",
    volume = "12",
    pages = "213",
    year = "2021"
}

@article{Anninos:2020hfj,
    author = "Anninos, Dionysios and Denef, Frederik and Law, Y. T. Albert and Sun, Zimo",
    title = "{Quantum de Sitter horizon entropy from quasicanonical bulk, edge, sphere and topological string partition functions}",
    eprint = "2009.12464",
    archivePrefix = "arXiv",
    primaryClass = "hep-th",
    doi = "10.1007/JHEP01(2022)088",
    journal = "JHEP",
    volume = "01",
    pages = "088",
    year = "2022"
}

@article{Lewkowycz:2013nqa,
    author = "Lewkowycz, Aitor and Maldacena, Juan",
    title = "{Generalized gravitational entropy}",
    eprint = "1304.4926",
    archivePrefix = "arXiv",
    primaryClass = "hep-th",
    doi = "10.1007/JHEP08(2013)090",
    journal = "JHEP",
    volume = "08",
    pages = "090",
    year = "2013"
}

@article{Andreassen:2016cvx,
    author = "Andreassen, Anders and Farhi, David and Frost, William and Schwartz, Matthew D.",
    title = "{Precision decay rate calculations in quantum field theory}",
    eprint = "1604.06090",
    archivePrefix = "arXiv",
    primaryClass = "hep-th",
    doi = "10.1103/PhysRevD.95.085011",
    journal = "Phys. Rev. D",
    volume = "95",
    number = "8",
    pages = "085011",
    year = "2017"
}

@article{Iliesiu:2020qvm,
    author = "Iliesiu, Luca V. and Turiaci, Gustavo J.",
    title = "{The statistical mechanics of near-extremal black holes}",
    eprint = "2003.02860",
    archivePrefix = "arXiv",
    primaryClass = "hep-th",
    doi = "10.1007/JHEP05(2021)145",
    journal = "JHEP",
    volume = "05",
    pages = "145",
    year = "2021"
}

@article{Heydeman:2020hhw,
    author = "Heydeman, Matthew and Iliesiu, Luca V. and Turiaci, Gustavo J. and Zhao, Wenli",
    title = "{The statistical mechanics of near-BPS black holes}",
    eprint = "2011.01953",
    archivePrefix = "arXiv",
    primaryClass = "hep-th",
    reportNumber = "PUPT-2621",
    doi = "10.1088/1751-8121/ac3be9",
    journal = "J. Phys. A",
    volume = "55",
    number = "1",
    pages = "014004",
    year = "2022"
}

@article{Brown:2007sd,
    author = "Brown, Adam R. and Weinberg, Erick J.",
    title = "{Thermal derivation of the Coleman-De Luccia tunneling prescription}",
    eprint = "0706.1573",
    archivePrefix = "arXiv",
    primaryClass = "hep-th",
    reportNumber = "CU-TP-1178, KIAS-P07025",
    doi = "10.1103/PhysRevD.76.064003",
    journal = "Phys. Rev. D",
    volume = "76",
    pages = "064003",
    year = "2007"
}

@article{Banerjee:2023quv,
    author = "Banerjee, Nabamita and Saha, Muktajyoti",
    title = "{Revisiting leading quantum corrections to near extremal black hole thermodynamics}",
    eprint = "2303.12415",
    archivePrefix = "arXiv",
    primaryClass = "hep-th",
    doi = "10.1007/JHEP07(2023)010",
    journal = "JHEP",
    volume = "07",
    pages = "010",
    year = "2023"
}

@article{Kapec:2023ruw,
    author = "Kapec, Daniel and Sheta, Ahmed and Strominger, Andrew and Toldo, Chiara",
    title = "{Logarithmic Corrections to Kerr Thermodynamics}",
    eprint = "2310.00848",
    archivePrefix = "arXiv",
    primaryClass = "hep-th",
    doi = "10.1103/PhysRevLett.133.021601",
    journal = "Phys. Rev. Lett.",
    volume = "133",
    number = "2",
    pages = "021601",
    year = "2024"
}

@article{Blacker:2025zca,
    author = "Blacker, Matthew J. and Castro, Alejandra and Sybesma, Watse and Toldo, Chiara",
    title = "{Quantum corrections to the path integral of near extremal de Sitter black holes}",
    eprint = "2503.14623",
    archivePrefix = "arXiv",
    primaryClass = "hep-th",
    reportNumber = "NORDITA 2025-016",
    doi = "10.1007/JHEP08(2025)120",
    journal = "JHEP",
    volume = "08",
    pages = "120",
    year = "2025"
}

@article{Maldacena:1997re,
    author = "Maldacena, Juan Martin",
    title = "{The Large $N$ limit of superconformal field theories and supergravity}",
    eprint = "hep-th/9711200",
    archivePrefix = "arXiv",
    reportNumber = "HUTP-97-A097, HUTP-98-A097",
    doi = "10.4310/ATMP.1998.v2.n2.a1",
    journal = "Adv. Theor. Math. Phys.",
    volume = "2",
    pages = "231--252",
    year = "1998"
}

@article{Gubser:1998bc,
    author = "Gubser, S. S. and Klebanov, Igor R. and Polyakov, Alexander M.",
    title = "{Gauge theory correlators from noncritical string theory}",
    eprint = "hep-th/9802109",
    archivePrefix = "arXiv",
    reportNumber = "PUPT-1767",
    doi = "10.1016/S0370-2693(98)00377-3",
    journal = "Phys. Lett. B",
    volume = "428",
    pages = "105--114",
    year = "1998"
}

@article{Witten:1998qj,
    author = "Witten, Edward",
    title = "{Anti de Sitter space and holography}",
    eprint = "hep-th/9802150",
    archivePrefix = "arXiv",
    reportNumber = "IASSNS-HEP-98-15",
    doi = "10.4310/ATMP.1998.v2.n2.a2",
    journal = "Adv. Theor. Math. Phys.",
    volume = "2",
    pages = "253--291",
    year = "1998"
}

@article{Almheiri:2019qdq,
    author = "Almheiri, Ahmed and Hartman, Thomas and Maldacena, Juan and Shaghoulian, Edgar and Tajdini, Amirhossein",
    title = "{Replica Wormholes and the Entropy of Hawking Radiation}",
    eprint = "1911.12333",
    archivePrefix = "arXiv",
    primaryClass = "hep-th",
    doi = "10.1007/JHEP05(2020)013",
    journal = "JHEP",
    volume = "05",
    pages = "013",
    year = "2020"
}

@article{Penington:2019kki,
    author = "Penington, Geoff and Shenker, Stephen H. and Stanford, Douglas and Yang, Zhenbin",
    title = "{Replica wormholes and the black hole interior}",
    eprint = "1911.11977",
    archivePrefix = "arXiv",
    primaryClass = "hep-th",
    doi = "10.1007/JHEP03(2022)205",
    journal = "JHEP",
    volume = "03",
    pages = "205",
    year = "2022"
}

@book{Polchinski:1998rq,
    author = "Polchinski, J.",
    title = "{String theory. Vol. 1: An introduction to the bosonic string}",
    doi = "10.1017/CBO9780511816079",
    isbn = "978-0-511-25227-3, 978-0-521-67227-6, 978-0-521-63303-1",
    publisher = "Cambridge University Press",
    series = "Cambridge Monographs on Mathematical Physics",
    month = "12",
    year = "2007"
}

@article{Horowitz:2025zpx,
    author = "Horowitz, Gary T. and Marolf, Donald and Santos, Jorge E.",
    title = "{Constraints are not enough}",
    eprint = "2505.13600",
    archivePrefix = "arXiv",
    primaryClass = "hep-th",
    month = "5",
    year = "2025"
}

@article{Chandrasekaran:2022cip,
    author = "Chandrasekaran, Venkatesa and Longo, Roberto and Penington, Geoff and Witten, Edward",
    title = "{An algebra of observables for de Sitter space}",
    eprint = "2206.10780",
    archivePrefix = "arXiv",
    primaryClass = "hep-th",
    doi = "10.1007/JHEP02(2023)082",
    journal = "JHEP",
    volume = "02",
    pages = "082",
    year = "2023"
}

@article{Witten:2023xze,
    author = "Witten, Edward",
    title = "{A background-independent algebra in quantum gravity}",
    eprint = "2308.03663",
    archivePrefix = "arXiv",
    primaryClass = "hep-th",
    doi = "10.1007/JHEP03(2024)077",
    journal = "JHEP",
    volume = "03",
    pages = "077",
    year = "2024"
}

@article{Chen:2025jqm,
    author = "Chen, Yiming and Stanford, Douglas and Tang, Haifeng and Yang, Zhenbin",
    title = "{On the phase of the de Sitter density of states}",
    eprint = "2511.01400",
    archivePrefix = "arXiv",
    primaryClass = "hep-th",
    doi = "10.1007/JHEP05(2026)068",
    journal = "JHEP",
    volume = "05",
    pages = "068",
    year = "2026"
}

@article{Ivo:2025xek,
    author = "Ivo, Victor and Sun, Zimo",
    title = "{The phase of charged Nariai solutions}",
    eprint = "2511.06604",
    archivePrefix = "arXiv",
    primaryClass = "hep-th",
    month = "11",
    year = "2025"
}

@article{Giombi:2026sqa,
    author = "Giombi, Simone and Sun, Zimo",
    title = "{The phase of de Sitter higher spin gravity}",
    eprint = "2601.15257",
    archivePrefix = "arXiv",
    primaryClass = "hep-th",
    month = "1",
    year = "2026"
}

@article{Ivo:2026ijv,
    author = "Ivo, Victor and Tang, Haifeng",
    title = "{One-loop aspects of de Sitter axion wormholes}",
    eprint = "2603.02335",
    archivePrefix = "arXiv",
    primaryClass = "hep-th",
    month = "3",
    year = "2026"
}

@article{Shi:2026wnk,
    author = "Shi, Xiaoyi and Turiaci, Gustavo J. and Wu, Chih-Hung",
    title = "{The fate of nucleated black holes in de Sitter quantum gravity}",
    eprint = "2605.03015",
    archivePrefix = "arXiv",
    primaryClass = "hep-th",
    doi = "10.1007/JHEP08(2026)191",
    journal = "JHEP",
    volume = "08",
    pages = "191",
    year = "2026"
}

@article{Ookouchi:2026bqj,
    author = "Ookouchi, Yutaka",
    title = "{The Origin and Fate of Rapid Negative Modes in Coleman--De Luccia Tunneling}",
    eprint = "2609.37092",
    archivePrefix = "arXiv",
    primaryClass = "hep-th",
    month = "9",
    year = "2026"
}
\end{document}